# Verification of high-energy transport codes on the basis of activation data

Yu. E. Titarenko[1], V. F. Batyaev[1], M. A. Butko[1], D. V. Dikarev[1], S. N. Florya[1], K. V. Pavlov[1], A. Yu. Titarenko[1], R. S. Tikhonov[1], V. M. Zhivun[1], A. V. Ignatyuk[2], S. G. Mashnik[3], A. Boudard[4], S. Leray[4], J.-C. David[4], J. Cugnon[5], D. Mancusi[5], Y. Yariv[6], H. Kumawat[7], K. Nishihara[8], N. Matsuda[8], G. Mank[9], W. Gudowski[10]

[1] Institute for Theoretical and Experimental Physics, 117218 Moscow, Russia

[2] Institute of Physics and Power Engineering, 249033 Obninsk, Russia

[3] Los Alamos National Laboratory, NM 87545, USA

[4] CEA, Saclay, 91191 Gif-sur-Yvette CEDEX, France

[5] University of Liege, 4000 Liege, Belgium

[6] Soreq Nuclear Research Center, 81800 Yavne, Israel

[7] Bhabha Atomic Research Center, 400085 Mumbai, India

[8] Japan Atomic Energy Agency, Ibaraki 319-1195, Japan

[9] International Atomic Energy Agency, A-1400 Vienna, Austria

[10] Royal Institute of Technology, 10691 Stockholm, Sweden

Nuclide production cross sections measured at ITEP for the targets of $^{nat}$Cr, $^{56}$Fe, $^{nat}$Ni, $^{93}$Nb, $^{181}$Ta, $^{nat}$W, $^{nat}$Pb, $^{209}$Bi irradiated by protons with energies from 40 to 2600 MeV were used to estimate the predictive accuracy of several popular high-energy transport codes. A general agreement of the ITEP data with the data obtained by other groups, including the numerous GSI data measured by the inverse kinematics method was found. Simulations of the measured data were performed with the MCNPX (Bertini and ISABEL options), CEM03.02, INCL4.2+ABLA, INCL4.5+ABLA07, PHITS, and CASCADE.07 codes. Deviation factors between the calculated and experimental cross sections have been estimated for each target and for the whole energy range covered by our measurements. Two-dimensional diagrams of deviation factor values were produced for estimating the predictive power of every code for intermediate, not measured masses of nuclei-targets and bombarding energies of protons. Further improvements of all tested here codes are recommended.

In addition, new measurements at ITEP of nuclide yields from a $^{208}$Pb target irradiated by 500 MeV protons are presented. A good agreement between these new data and the GSI measurements obtained by the inverse kinematics method was found.

## I. INTRODUCTION

High-energy transport codes (HETC) based on various versions of nuclear reaction models (generally, an intranuclear cascade model (INC) followed by different de-excitation models) are widely used in many nuclear centers for the analysis of experimental data, for calculations of nuclear accelerator shielding, and for design of new nuclear facilities. Each of the available codes was developed originally for well specified tasks and the model parameters were estimated on the basis of fitting the corresponding data. With an expansion of tasks, the agreement between predictions by different HETC diminishes, as a rule; examples of which are presented in many publications [1-3].

HETC are used mainly for calculations of two types of data: 1) experiments with thin targets* aimed at studying the initial processes of the incident particle interaction with the target nuclei; 2) simulations of thick targets, in which an essential role is played by the secondary processes induced by particles from the primary interactions, involving the internuclear transport of all produced particles. Obviously, the second type of data is important for most of the practical applications. Our present work is restricted to only the first type of data, providing more information about the initial intranuclear processes.

In connection with the widely discussed designs of accelerator driven power systems, a significant amount of experimental data on the proton induced reactions was accumulated in the last 10-15 years for a broad range of proton energies and target masses [4]. These data were partially analyzed in the original works with different versions of HETC. Recently, a benchmark of spallation models has been organized under the auspices of IAEA on a selected set of experimental data covering different reaction channels [3]. In the present work, we focus on a consistent set of the most accurate experimental data concerning yields of residual nuclei and try to estimate the predictive powers of several most popular HETC.

The available data on the independent and cumulative product yields of the proton-induced reactions are briefly discussed in the first section. In the second section, the overlapping data measured at different laboratories are compared and consistent data-sets are selected, which are used later for comparison with the HETC results. The main components and parameters of HETC are briefly described in the third section. The fours section presents a comparison of the model calculations with experimental data. For each code, deviation factors between the model simulations and experiments are determined for all considered targets and proton energies from 40 up to 2600 MeV. Contour plots of the deviation factor values are produced to estimate the predictive power of each code for intermediate proton energies and

* A thin target must satisfy two criteria: 1) the incident particle energy loss in the target is negligible compared with the initial energy, and 2) the incident particle free path is much larger than the target thickness. In other cases, the target is considered thick.

masses of targets not covered by our current measurements. The main conclusions together with a recommendation to further improve all tested here codes are presented in the last section.

## II. EXPERIMENTAL DATA ON INDEPENDENT AND CUMULATIVE YIELDS OF REACTION PRODUCTS

Nowadays, the residual yields of nuclear reaction products are measured by means of two methods which are usually named “direct” and “inverse kinematics”:

- In the framework of the first one, a thin target is irradiated by a proton beam ($^{1}_{1}H \rightarrow ^{A}_{z}T$) and reaction products are measured after chemical separation, or without it, by means of α-, β-, or γ-spectrometers including high-resolution silicon, germanium, or scintillating detectors [4-22].
- In case of the inverse kinematics, a liquid hydrogen target is irradiated by the corresponding heavy-ion beam ($^{A}_{z}T \rightarrow ^{1}_{1}H$) and the charge and mass distributions of reaction products are measured by means of a high-resolution magnetic separator and a ionization chamber [23-28].

For the direct method, named frequently as the “activation” one, the off-line regime of measurements is used, as a rule, and only the radioactive reaction products are detected. Radioactive nuclides can be produced in both the ground and isomeric states, whose yields are measured separately. Such nuclides are produced not only in the initial nuclear reaction, but also after a decay of its chain precursors. Accordingly, the cumulative yields of nuclides are measured in the most cases.

With the inverse kinematics method, the reaction products are measured in the time intervals about $10^{-7}$- $10^{-9}$ s after the nuclear reaction. So, the independent yields of residual nuclei are registered without any separation of the ground and isomeric states.

In accordance with the standard terminology, we will use below the symbols **i** and **c** for the independent and cumulative yields, respectively, the additional symbols **(g)** and **(m)** for the identification of the ground and metastable isomeric states, and the symbol **c*** for the supra-cumulative yields, whose difference from the cumulative ones is discussed below. All types of product yields are measured by the activation methods while only independent ones are obtained in the inverse kinematics approach.

Only the independent yields are calculated by all HETC in this work. Obviously, such yields are more convenient for verification of the models included in the codes. However, for practical applications, the main requests on the data are connected with the long-lived nuclides which determine a residual radioactivity of irradiated components at nuclear facilities. Such activities are connected mainly with the cumulative yields.

Nowadays, the decay schemes are well known for most of the radioactive nuclei and a transition from the independent to the cumulative yields can be calculated rather accurately. As an example, we shall consider the chain of nuclides connected through the $\beta^-$, ε, α and other possible transitions. All nuclides in the chain can be numerated independently from the atomic number in such a way that a nuclide with a smaller number decays to a nuclide with a larger number [14]. Then, for the established cross sections of independent product yields the cumulative production cross section can be calculated as

$$s_k^{cum} = \sum_{j=1}^{k} m_{kj} \cdot s_j^{ind}, \qquad k = 1, 2, ..., n \ . \qquad (1)$$

Here $m_{kj}$ is the matrix whose elements are determined in accordance with the formulas

$$m_{kj} = \begin{cases} \sum_{i=j}^{k-1} n_{ik} \cdot m_{ij} \ , & for \quad \kappa > j \\ 1 \ , & for \quad \kappa = j \\ 0 \ , & for \quad \kappa < j \end{cases} \qquad (2)$$

where $n_{ik}$ are the branching coefficients determining the probabilities of the various decay modes. Obviously, the cumulative cross section coincides with the individual one if for some nuclide there is no decaying precursor.

More detailed consideration of the above relations is presented in Refs. [6,7,14,16]. It should be noted, that in some cases the measured decay activities do not allow to separate completely the contributions of the mother and daughter nuclides. Such cases arise frequently when the time between the irradiation and the beginning of activity measurements is large compared to the half-life of the mother nuclide. The ZSR group has named such cross sections as "cumulative" too and used for their calculations the relation [22]:

$$s_{D,cum} = s_{D,ind} + s_{M,cum/ind} \frac{l_M}{l_M - l_D} \quad , \qquad (3)$$

where $l$ is the corresponding decay constant, inverse of the half-life of the mothers and daughter nuclides, respectively.

The ITEP group has proposed to name such cross sections as the "supra-cumulative" (cum*). Taking into account the branching coefficients, the supra-cumulative cross sections can be defined instead of using Eq. (3) by the formula:

$$s_2^{cum*} = s_2^{ind} + \frac{l_1}{l_1 - l_2} \cdot n_{1,2} \cdot s_1^{cum/ind} = s_2^{cum} + \frac{l_2}{l_1 - l_2} \cdot n_{1,2} \cdot s_1^{cum/ind} . \qquad (4)$$

The supra-cumulative cross section is always larger than the cumulative one, because $l_1 > l_2$. One can see from Eq. (4) that a difference between the cumulative and supra-cumulative cross sections can be rather large in the case of a small difference between the decay constants of the mother (index "1") and daughter (index "2") nuclei. More detailed considerations about the differences between the “cumulative” and “supra-cumulative” cross sections are presented in Refs. [6,7,14,16]. These differences should certainly be taken into account in the model simulations of experimental data.

It is obvious from Eqs. (1) and (2) that the accuracy of the cumulative cross-section calculations depends essentially on completeness of our knowledge about the decay chains and branching coefficients. Such data were selected from the current version of the ENSDF library [29] that includes up to 18 decay modes, namely $\beta^-$, $\beta^- n$, $IT$, $\varepsilon$, $\varepsilon + \beta^+$, $p$, $\alpha$, $\beta^+ p$, $\beta^+ \alpha$, $\beta^+ 2p$, $\varepsilon p$, $\varepsilon\alpha$, $2\varepsilon$, $n$, $\beta^+$, $2\beta^-$, $2\beta^+$, 2e. The modes leading to the same changes of the mass and charge numbers were combined into 12 decay groups. The branching coefficients were taken for the ground states or for the first metastable states, if there was no the ground state data. To construct the matrixes of all possible transitions (2), a special code was developed that considers the following approximations:

- if only the limit value of a branching factor is given in ENSDF, the factor value was taken as equal to that limit;
- if a branching coefficient is indicated as unknown (the symbol “?” is shown), its value was taken equal to zero;
- if several decay modes are presented, but the sum of their values is below 100%, the unknown branching factor value was taken to be the difference between 100% and the sum of the remaining decay modes;
- if but a single decay mode with a branching factor below 100% is presented, the missing difference was ascribed to either mode $\beta^+$ (for neutron-deficient nuclides) or mode $\beta^-$ (for neutron-rich nuclides);
- if a delayed decay mode is presented, the branching factor of this mode was always subtracted from the factor value of the corresponding dominant mode;

It should be noted that the branching factor values taken as described above in the ENSDF file are sometimes very different from the values of other sources. For instance, $\alpha \leq 100$ %, $\varepsilon + \beta^+$-? is indicated in ENSDF for $^{177}$Au, whereas the Nuclear Wallet Cards (April 2005) gives $\varepsilon \geq 60$ % [30].

For the reasons mentioned above and considering that such information is important when verifying the codes, the Technical reports on Projects # 2002 and # 3266 present all chains together with their branching factors. Fig. 1 shows an example of the $^{148}$Gd chain that includes 47 precursors.
For convenience, the analyzed decay chains were broken into the chains of $^{209}$Bi, $^{nat}$Pb, $^{nat}$W, $^{181}$Ta, $^{93}$Nb, $^{nat}$Ni, $^{56}$Fe and $^{nat}$Cr decays.

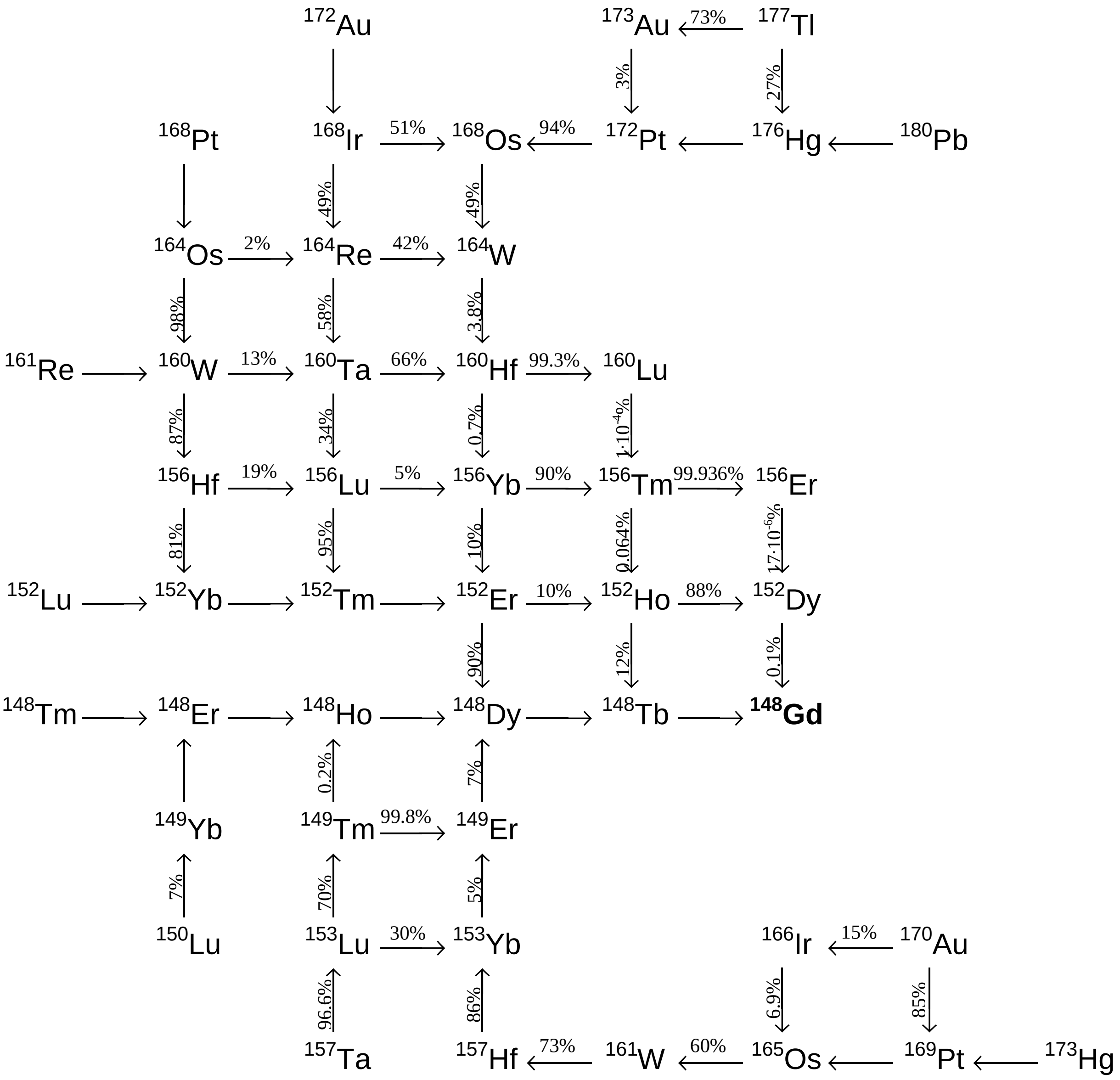

Fig. 1. Chain of decays for the production of $^{148}$Gd.

### III. SELECTION OF EXPERIMENTAL DATA

Experimental data used for verification of HETC should satisfy the following requirements:

- the number of irradiated targets should be significant;

- the data should be obtained for a wide energy interval.

At present, such requirements can be satisfied the best way if we considered jointly the recent data obtained at GSI (Darmstadt), ZSR (Hannover), and at ITEP (Moscow) [4-28]. The combined analysis of all these data allows us to estimate their consistency and virtual accuracy.

The total number of the production cross sections measured at ITEP during 1997-2010 is equal to 14621. The list of the irradiated targets and incident proton energies is presented in Table I. Each cell of this table indicates the number of the residual nuclides measured for the corresponding target and the proton incident energy [4-21].

TABLE I. List of targets irradiated at the ITEP U-10 accelerator and the number of the measured production cross sections for every bombarding proton energy.

| Proton energy (GeV) | Targets | | | | | | | | | | | | | | | | | | | | | |
|---|---|---|---|---|---|---|---|---|---|---|---|---|---|---|---|---|---|---|---|---|---|---|
| | $^{nat}$Cr | $^{56}$Fe * | $^{nat}$Ni | $^{59}$Co | $^{63}$Cu | $^{65}$Cu | $^{93}$Nb | $^{99}$Tc | $^{181}$Ta | $^{182}$W | $^{183}$W | $^{184}$W | $^{186}$W | $^{nat}$W | $^{197}$Au | $^{nat}$Hg | $^{206}$Pb | $^{207}$Pb | $^{208}$Pb | $^{nat}$Pb | $^{209}$Bi | $^{232}$Th | $^{nat}$U |
| 0.04 | 15 | 17 | 20 | | | | 19 | | 9 | | | | | 19 | | | 13 | 9 | 8 | 18 | 13 | | |
| 0.07 | 17 | 21 | 22 | | | | 28 | | 17 | | | | | 31 | | | 28 | 29 | 28 | 28 | 35 | | |
| 0.1 | 19 | 24 | 27 | | | | 37 | 18 | 31 | | | | | 44 | | 44 | 46 | 42 | 36 | 43 | 50 | 87 | 108 |
| 0.13 | | | | 24 | 11 | 6 | | | | | | | | | | | 22 | 22 | 20 | | 26 | | |
| 0.15 | 25 | 25 | 28 | | | | 46 | | 40 | | | | | 53 | | | 65 | 65 | 63 | 63 | 71 | | |
| 0.2 | | | | 29 | 29 | 29 | | 39 | | 32 | 35 | 36 | 36 | | | 65 | | | | | | 128 | 123 |
| 0.25 | 28 | 33 | 37 | | | | 58 | | 53 | | | | | 69 | | | 94 | 94 | 94 | 95 | 106 | | |
| 0.3 | | 36 | | | | | | | | | | | | | | | | | | | | | |
| 0.4 | 33 | 36 | 36 | | | | 65 | | 81 | | | | | 82 | | | 112 | 112 | 113 | 116 | 128 | | |
| 0.5 | | 33 | | | | | | | | | | | | | | | | | 128* | | | | |
| 0.6 | 33 | 38 | 40 | | | | 76 | | 99 | | | | | 105 | | | 139 | 140 | 141 | 141 | 147 | | |
| 0.75 | | 38 | | | | | | | | | | | | | | | | | | | | | |
| 0.8* | 33 | 38 | 43 | | | | 86 | 72 | 100 | 70 | 76 | 77 | 62 | 113 | 101 | 103 | 156 | 152 | 154 | 154 | 162 | 130 | 195 |
| 1.0 | | 38 | | | | | | 64 | | | | | | | | | | | 114 | | | | |
| 1.2 | 33 | 39 | 43 | 41 | 47 | 54 | 96 | 67 | 143 | | | | | 155 | | | 170 | 170 | 170 | 171 | 183 | 214 | 226 |
| 1.5 | | 38 | | | 35 | 36 | | | | | | | | | | | 92 | 93 | 94 | 93 | 99 | | |
| 1.6 | 33 | 38 | 46 | 41 | 42 | 47 | 106 | 78 | 146 | 109 | 111 | 114 | 119 | 163 | | | 180 | 180 | 182 | 181 | 192 | 212 | 231 |
| 2.6 | 33 | 38 | 46 | 41 | 42 | 48 | 107 | 85 | 163 | | | | | 179 | | 141 | 171 | 171 | 172 | 178 | 198 | | |

To test the reliability of the ITEP data, they were compared with similar data obtained by other groups. The following data were considered:

- production cross sections for residual radioactive nuclides on targets of $^{63}$Cu and $^{65}$Cu irradiated in pairs by 1.2 GeV protons at the ITEP accelerator and analyzed independently with spectrometers at ITEP and JAERI (Tokai) [6];

- nuclide production cross sections from $^{56}$Fe obtained by the direct and inverse kinematics methods at ITEP and GSI, respectively, for the corresponding proton energies of 0.3, 0.5, 0.75, 1.0, and 1.5 GeV [14,23];
- nuclide production cross sections from $^{197}$Au obtained by the activation method at ITEP and ZRS and by the inverse kinematics method at GSI for the proton energy of 0.8 GeV [8,24];
- nuclide production cross sections from $^{208}$Pb obtained by the direct and inverse kinematics method at ITEP and GSI, respectively, for the proton energies of 0.5 and 1.0 GeV [6-8,25-27]. It should be noted that the ITEP activation data at 0.5 GeV were measured specifically for our comparison of the two methods and they are presented here in the Appendix;
- nuclide production cross sections from the targets of $^{nat}$Cr, $^{56}$Fe, $^{nat}$Ni, $^{93}$Nb, $^{181}$Ta, $^{nat}$W, $^{nat}$Pb, $^{209}$Bi obtained by the activation method at ITEP and ZSR for the proton energies of 0.4 – 2.6 GeV [4,18-22].

Here, we use the mean-square deviation factor as criteria of a general consistency between different experimental data, defined as:

$$F = 10^{\left(\frac{1}{N}\sum_{i=1}^{N}[\log(s_i^{ITEP}/s_i^{other})]^2\right)^{1/2}}, \qquad (5)$$

where the superscript of the cross-section symbol indicates the source of data and N is the number of the points included in the comparison (see, e.g., Ref. [31]).

Table II presents the calculated deviation factors between the data of ITEP and other institutes, at energies above 300 MeV. Table III shows a specific comparison between the data obtained by ITEP and ZSR.

The deviation factors calculated for the complete set of the available data are marked with the subscript *c*, and the deviation factors for the reduced set, in which the deviations exceeding three experimental uncertainties were rejected, are marked with the subscript 3σ. Each cell in the tables shows the total number of experimental points $N_0$ measured by the corresponding group, the number of points $N_c$ included in the calculations of the deviation factor $F_c$ , and the similar information for the reduced set of experimental points, $N_{3\sigma}$ and $F_{3\sigma}$, respectively. The numbers of points are given in the upper half of each cell and the corresponding deviation factors are presented in the lower half.

The deviation factors presented in Tables II and III allow us to draw the following conclusions:

1) The smallest deviations ($F_c$ = 1.11 and 1.14) display the ITEP-JAERI data sets obtained for the same irradiation conditions. For the ITEP-ZSR data sets obtained with different accelerators and for different targets (enriched isotopes $^{63}$Cu and $^{65}$Cu were irradiated at ITEP and a $^{nat}$Cu target was irradiated at ZSR), the deviation factors are only a little worse ($F_c$ = 1.15).

2) An important advantage of the inverse kinematics method is a large amount of measured data ( $N_0^{GSI}$ = 2947, for 8 experiments). The amount of the analogous data for the direct method is smaller by a factor of 6.5. For $^{56}$Fe, the deviation factors are similar to the values presented above for Cu ($F_c$ = 1.23 and $F_{3\sigma}$= 1.15) at high energies, but they increase at lower energies.

3) For heavier targets of gold and lead, the deviation factors for the ITEP-GSI data sets are larger: $F_c$ = 1.31-1.57 and $F_{3\sigma}$ = 1.27-1.29. The available data for lead display an increase of the deviation factors at lower energies, similar to the situation we discussed above for iron.

TABLE II. Average deviation factors between data sets measured at different institutes.

| Nuclides | Institu-tions | $N_0^{ITEP} / N_0^{JERI,ZSR,GSI} / N_c / N_{3s}$<br>$\langle F_c \rangle / \langle F_{3s} \rangle$ | | | | | | |
|---|---|---|---|---|---|---|---|---|
| | | $E_p$=300 MeV | 500 MeV | 750 MeV | 800 MeV | 1000 MeV | 1200 MeV | 1500 MeV |
| $^{63}$Cu | ITEP – JAERI | – | – | – | – | – | 54/25/25/25<br>1.14/1.14 | – |
| $^{65}$Cu | ITEP – JAERI | – | – | – | – | – | 48/25/25/25<br>1.11/1.11 | – |
| $^{nat}$Cu | ITEP – ZSR | – | – | – | – | – | 54/21/21/21<br>1.15/1.15 | – |
| $^{56}$Fe | ITEP – GSI | 36/128/25/22<br>1.58/1.53 | 33/136/26/23<br>1.37/28 | 38/148/30/26<br>1.30/1.20 | – | 38/152/28/27<br>1.23/1.15 | – | 38/157/29/28<br>1.23/1.15 |
| $^{197}$Au | ITEP – GSI | – | – | – | 101/705/81/74<br>1.40/1.28 | – | – | – |
| $^{197}$Au | ZSR-GSI | – | – | – | 79/705/43/37<br>1.99/1.34 | – | – | – |
| $^{197}$Au | ITEP – ZSR | – | – | – | 101/79/37/34<br>1.30/1.26 | – | – | – |
| $^{208}$Pb | ITEP – GSI | – | 129/670/98/80<br>1.57/1.29 | – | – | 114/853/74/71<br>1.31/1.27 | – | – |
| $^{208}$Pb/$^{nat}$Pb | ZSR-GSI | – | – | – | – | 98/853/40/37<br>1.24/1.22 | – | – |
| $^{208}$Pb/$^{nat}$Pb | ITEP – ZSR – | – | – | – | – | 114/98/62/61<br>1.45/1.41 | – | – |

4) For the ZSR-GSI data sets, the deviation factors are rather large for gold: $F_c$ = 1.99 and $F_{3\sigma}$ = 1.34, but similar to the ITEP-GSI deviation factors for lead: $F_c$ = 1.24 and $F_{3\sigma}$ = 1.22. A direct comparison of the ITEP and ZSR data shows also some essential differences between these two experimental data sets: $F_c$ = 1.45 and $F_{3\sigma}$ = 1.41. Such relatively large deviation factors can be related to a difference in the irradiation conditions or/and to some systematic errors of the activation methods used at ITEP and ZSR.

As a whole, the performed analysis has shown that the average differences between the experimental data obtained by the GSI, ITEP, JAERI, and ZSR groups are of the order of only about 30%, that is much less than the divergence between the measured data and the theoretical calculations based on various versions of HETC, as will be shown in the following. A reasonable agreement between different experimental data sets allows us to limit the code testing to the ITEP data only. Their advantage over others data sets relate to a more broad energy range, a larger number of targets, and a larger amount of the measured radioactive nuclides, due to including more short-lived isotopes.

TABLE III. Average deviation factors between the data obtained at ITEP and ZSR.

| Energy MeV | $\frac{N_0^{ITEP} / N_0^{ZSR} / N_c / N_{3s}}{\langle F_c \rangle / \langle F_{3s} \rangle}$ | | | | | | | |
|---|---|---|---|---|---|---|---|---|
| | $^{nat}Cr$ | $^{56}Fe$ | $^{nat}Ni$ | $^{93}Nb$ | $^{181}Ta$ | $^{nat}W$ | $^{nat}Pb$ | $^{209}Bi$ |
| 40 | – | 17/11/9/6 | 20/13/12/11 | 19/8/7/5 | – | – | 18/26/11/9 | – |
| | | 3.98/1.35 | 1.25/1.16 | 1.46/1.15 | – | – | 2.00/1.41 | – |
| 70 | – | 21/14/11/6 | 22/12/11/11 | 28/12/10/10 | 17/8/6/3 | 31/10/8/7 | 28/50/17/14 | 35/27/18/14 |
| | | 2.69/1.30 | 1.30/1.30 | 1.38/1.38 | 2.31/1.17 | 1.21/1.21 | 1.53/1.45 | 1.37/1.22 |
| 100 | – | 24/18/13/11 | 27/17/16/16 | 37/11/9/9 | 31/18/13/11 | 44/16/11/10 | 43/58/24/22 | 50/57/26/26 |
| | | 1.60/1.18 | 1.25/1.25 | 1.20/1.20 | 1.89/1.25 | 1.27/1.17 | 1.68/1.49 | 1.21/1.21 |
| 150 | – | 25/28/16/15 | 28/18/17/17 | 46/15/13/13 | 40/23/18/16 | 53/19/15/14 | 63/70/33/32 | 71/76/37/35 |
| | | 1.34/1.24 | 1.15/1.15 | 1.15/1.15 | 1.45/1.40 | 1.18/1.11 | 1.26/1.25 | 1.32/1.23 |
| 250 | – | 33/17/11/9 | – | 58/17/16/13 | 53/29/22/21 | 69/27/22/22 | 95/99/52/51 | 106/87/50/47 |
| | | 1.40/1.17 | – | 1.33/1.27 | 1.57/1.43 | 1.37/1.37 | 1.21/1.19 | 1.29/1.23 |
| 400 | – | 36/21/14/13 | – | 65/23/20/20 | 81/39/33/28 | 82/32/25/20 | 116/84/60/54 | 128/73/47/38 |
| | | 2.98/1.17 | – | 1.18/1.18 | 1.70/1.35 | 1.67/1.36 | 1.43/1.28 | 1.42/1.26 |
| 600 | 33/10/10/10 | 38/23/13/10 | 40/6/6/6 | 76/30/26/26 | 99/58/44/42 | 105/43/38/35 | 141/99/63/62 | 147/78/64/60 |
| | 2.92/2.92 | 1.41/1.26 | 1.12/1.12 | 1.13/1.13 | 1.61/1.31 | 1.30/1.23 | 1.22/1.19 | 1.42/1.27 |
| 800 | – | 38/24/14/13 | 43/17/17/16 | 86/19/19/19 | 100/61/47/47 | 113/49/40/37 | 154/105/72/70 | 162/89/56/53 |
| | | 1.23/1.16 | 1.19/1.12 | 1.20/1.20 | 1.18/1.18 | 1.33/1.26 | 1.32/1.27 | 1.34/1.22 |
| 1200 | – | 39/28/18/14 | 43/22/22/21 | 96/35/33/33 | 143/63/55/54 | 155/51/47/45 | 171/105/75/73 | 183/102/72/68 |
| | | 4.85/1.12 | 1.27/1.22 | 1.15/1.15 | 1.24/1.22 | 1.36/1.31 | 1.26/1.22 | 1.22/1.19 |
| 1600 | – | 38/28/19/16 | 46/22/22/20 | 106/34/32/32 | 146/68/56/55 | 163/55/49/47 | 181/126/79/77 | 192/125/83/75 |
| | | 1.35/1.10 | 1.25/1.13 | 1.10/1.10 | 1.29/1.25 | 1.28/1.18 | 1.33/1.25 | 1.27/1.18 |
| 2600 | – | 38/31/19/16 | 46/22/22/21 | 107/25/25/24 | 163/86/73/70 | 179/56/55/50 | 178/132/89/86 | 198/108/76/71 |
| | | 2.47/1.63 | 1.45/1.31 | 1.64/1.21 | 1.39/1.34 | 1.40/1.26 | 1.33/1.27 | 1.37/1.27 |

## IV. CODES USED FOR CROSS-SECTION CALCULATIONS

The following five codes were used to simulate the measured cross section:

- MCNPX is a well-known transport code used extensively in various nuclear applications [32]. The code includes Monte-Carlo models simulating the intranuclear interactions of nucleons, pions, and other elementary particles and nuclei, and their subsequent transport in extended objects. The code has been developed at Los Alamos and it uses models of INC, pre-equilibrium emission, evaporation of nucleons, charged particles, and gammas, as well as several versions of the nuclear fission model. Both the Bertini and ISABEL versions of the INC model were used in the present calculations. The Dresner approach was used for the evaporation model, and the Atchison model (RAL) was applied for fission (see [32] for more details).
- CEM03.02 is the latest version of an improved Cascade-Exciton Model (CEM) [33], proposed initially at JINR (Dubna) [34]. Relatively to the previous CEM97 and CEM95 versions of CEM (see details and references in [33]), CEM03.03 has a longer cascade stage, a reduced pre-equilibrium emission, an improved photonuclear reaction model, and an evaporation stage with the higher initial excitation energy [35,36]. Besides, CEM03.02 includes an extended Fermi breakup model, production of energetic complex particles via coalescence of nucleons emitted during the INC stages of reactions, and a modification of the Generalized Evaporation (+fission) Model code GEM2 by Furihata [37]. CEM03.02 and its precursors have been incorporated in the MARS, MCNPX, and LAHET codes and were used in numerous applications. The current version is widely used at Los Alamos in MCNP6 and MCNPX transport codes for different fundamental and applied tasks.
- The INCL4.2+ABLA code is based on the Liege INC model [38] developed jointly with CEA (Saclay) combined with the GSI evaporation/fission model [39]. The codes have been developed under the HINDAS [40] (version INCL4.2+ABLA) and EUROTRANS/NUDATRA (version INCL4.5+ABLA07 [41,42]) projects, and is used extensively in many fundamental European projects. The INCL4.2+ABLA version has been implemented in the LAHET3 and MCNPX transport code-systems. The present calculations have been performed with both the previous INCL4.2+ABLA version and the more recent INCL4.5+ABLA07 version.
- The PHITS code is based on models similar to the ones included in the codes discussed above (see [43] and references therein). It has been used extensively in many applications of accelerator engineering, radiotherapy, and space exploration. PHITS includes three versions of the initial INC stage and the traditional model for evaporation and fission. The INC versions are the Bertini model [44,45], the **J**et AA Microscopic (JAM) transport model [46], and the JAERI Quantum-Molecular Dynamics (QMD) model [47]. The subsequent stage models for the evaporation and fission are adopted form the GEM code [37]. For the internuclear transport calculations, all neutron cross sections can be taken from the international libraries of evaluated data.

- CASCADE07 is an upgrade of the CASCADE code [49-52] developed at JINR, Dubna to simulate reactions on both thin and thick targets. The INC is described with an old version of the Dubna intranuclear cascade model able to calculate nucleus-nucleus interactions (often referred in the literature simply as the Dubna Cascade Model (DCM) [53]). The hadron-nucleus interaction cross sections are calculated in accordance with the compiled experimental data [54] and the nucleus-nucleus cross-sections are calculated via analytical approximations with the parameters determined in [55]. Special attention was spent to an adjustment of the cutoff energy determining the end of INC stage. The pre-equilibrium stage is based on the Modified-Exciton Model (MEM) [56]. The transition to the equilibrium evaporation stage has been modified according to [57]. The evaporation and fission processes are treated in accordance with the description presented in [58].

Let us note that a brief description of all codes used here together with many useful references may be found on-line in the "Ingredients" column of the **Calculation Results** Section of the materials on the recent benchmark of spallation models [3].

The experimental data obtained in Refs. [12,16,18-21] were simulated at 35 proton energies ranging from 0.033 to 3.5 GeV in order to get smooth excitation functions for each residual nuclide. Calculations with the Bertini, ISABEL, INCL4.2+ABLA, INCL4.5+ABLA07 codes were performed at ITEP, and similar calculations with the CEM03.02, PHITS, CASCADE07 codes were carried out by the code authors directly at their laboratories.

All calculated excitation functions were plotted and compared with the ITEP experimental data and available data by other laboratories. The number of such plots is equal to 1467 and the visual control is important to remove possible accidental errors related to both the experimental data processing and the fluctuations of calculated results owing to low statistics at near threshold energies. The plots are also useful for comparison of experimental data obtained at different energies, where a direct comparison can be incorrect. Partially, the plots were published in Refs. [12,16,18-21].

Beside the excitation functions, distributions of the ratios $\sigma_{exp} / \sigma_{calc}$ were analyzed, too. 198 plots were prepared and their visual control helps in removing some possible accidental errors in both experimental and theoretical results.

Each of the above codes makes use of its own value for the total reaction cross sections. To get a correct comparison among the excitation functions obtained by different codes, the calculated results were renormalized to a single reaction cross section value obtained with the Letaw formula [59].

As an example, the results of calculated mass distributions of 800 MeV proton-induced reaction products are compared in Fig. 2 with the measured cumulative and supra-cumulative yields for the targets $^{93}$Nb and $^{nat}$W. Since the cumulative yields correspond to but a fraction of the products, their difference from the calculated ones characterizes the contributions from the produced stable isotopes and radioactive isotopes that do not belong to the respective beta-decay chains.

All codes provide a sufficiently good description of the measured mass yields of the products close to target nucleus mass ($A_{tar}$-$A$) < 25–30). In the region of low masses ($A$ < 40), however, a reasonable description of the observed yields is only obtained with the codes that, apart from the conventional evaporation of light particles, allow for evaporation of heavy clusters (the INCL4.5+ABLA07, CEM03.02, and PHITS codes). For heavy targets, an accurate simulation of nuclear fission is important. The results for $^{nat}$W demonstrate that such simulation for the INCL4.5+ABLA07 and CASCADE07 codes is still far from optimal. Divergences between the calculated and experimental data similar to the one shown in Fig. 2 may be observed also for others proton energies and other targets. Therefore, we can conclude that none of the codes tested here provide a good quantitative description of the whole set of measured data.

## V. PREDICTIVE ACCURACIES OF HETC

For a quantitative analysis of differences between the experimental data and calculations, we have to use the factor of mean-square deviations determined in accordance with Eq. (5), in which the *other* data are replaced by the *calculated* ones. Calculations were carried out with all the above mentioned codes for the proton energies coinciding with the experimental ones. The number of cross sections included in the analyses for each target and the corresponding energy is presented in Table IV.

Each cell of Table IV contains the number of the measured nuclide yields and the number of yields included in calculations of the deviation factors. The difference between these numbers is due to excluding from calculations of the deviation factors the data for meta-stable states and the cases when the calculated yields were obtained with too low statistics (less than several events from about 5-10 million of simulated events).

The deviation factors obtained for each code at the corresponding energy are given in Table V. The lowest values of factors for every energy are marked by bold front and the largest ones are marked by underlined.

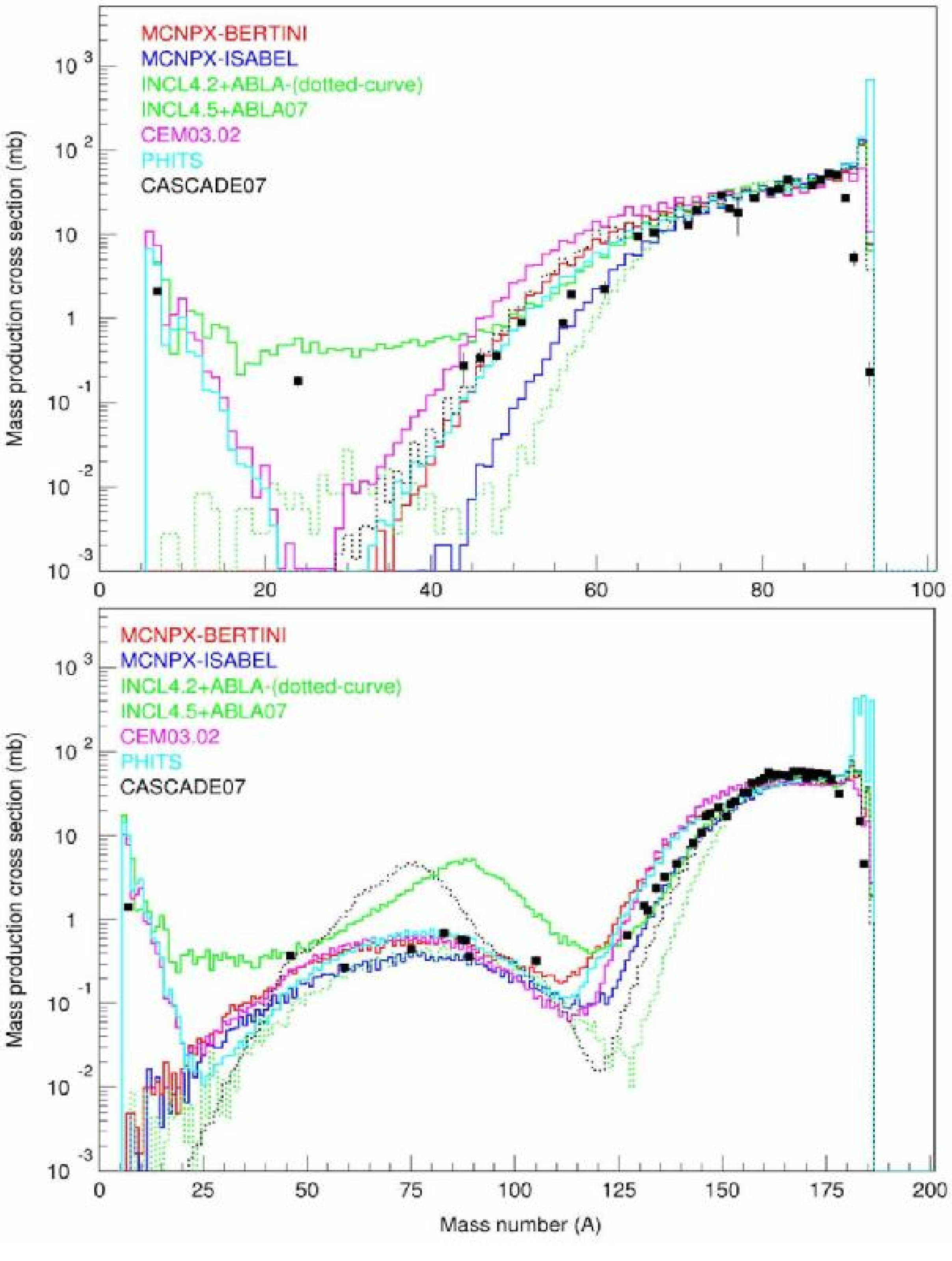

Fig. 2. Calculated mass distributions of product yields for the targets of $^{93}$Nb (upper plot) and $^{nat}$W (lower plot) irradiated with 0.8-GeV protons compared with the measured cumulative and supra-cumulative yields.

TABLE IV. Total number of the production cross sections measured for the indicated targets and proton energies and the portion of them used for comparisons with the calculated cross sections.

| Tar-gets | Proton energy (GeV) | | | | | | | | | | | Total |
|---|---|---|---|---|---|---|---|---|---|---|---|---|
| | 0.04 | 0.07 | 0.1 | 0.15 | 0.25 | 0.4 | 0.6 | 0.8 | 1.2 | 1.6 | 2.6 | |
| $^{209}$Bi | 13/12 | 35/32 | 50/45 | 71/58 | 106/81 | 128/98 | 147/111 | 162/125 | 183/147 | 192/158 | 198/162 | 1285/103 |
| $^{nat}$Pb | 18/15 | 28/23 | 43/33 | 63/42 | 95/69 | 116/88 | 141/106 | 154/119 | 171/135 | 181/145 | 178/148 | 1188/923 |
| $^{nat}$W | 19/16 | 31/27 | 44/38 | 53/47 | 69/64 | 82/75 | 105/93 | 113/97 | 155/131 | 163/133 | 179/144 | 1013/865 |
| $^{nat}$Ta | 9/5 | 17/13 | 31/28 | 40/36 | 53/45 | 81/71 | 99/88 | 100/90 | 143/117 | 146/121 | 163/133 | 882/747 |
| $^{93}$Nb | 19/12 | 28/16 | 37/20 | 46/29 | 58/38 | 65/42 | 76/52 | 86/62 | 96/70 | 106/81 | 107/82 | 724/504 |
| $^{nat}$Ni | 20/18 | 22/20 | 27/24 | 28/25 | 37/32 | 36/31 | 40/35 | 43/37 | 43/37 | 46/40 | 46/40 | 388/339 |
| $^{56}$Fe | 17/12 | 21/15 | 24/19 | 25/20 | 33/54 | 36/59 | 38/64 | 38/64 | 39/63 | 38/31 | 38/31 | 347/432 |
| $^{nat}$Cr | 15/12 | 17/15 | 19/17 | 25/18 | 28/24 | 33/26 | 33/29 | 33/29 | 33/29 | 33/29 | 33/29 | 302/257 |
| Total | | | | | | | | | | | | 6129/509 |

TABLE V. Estimated deviation factors between the experimental and simulated data for every code.

| Codes | Tar-gets | Proton energy (GeV) | | | | | | | | | | | |
|---|---|---|---|---|---|---|---|---|---|---|---|---|---|
| | | 0.04 | 0.07 | 0.1 | 0.15 | 0.25 | 0.4 | 0.6 | 0.8 | 1.2 | 1.6 | 2.6 | Mean |
| BERTINI | Cr | 3.54 | 3.06 | 3.27 | 2.42 | **1.86** | 4.39 | 2.82 | 4.14 | 2.96 | 2.63 | 2.26 | 3.02 |
| | Fe | 5.86 | 3.06 | 3.81 | 2.64 | 2.98 | 3.47 | 2.77 | **2.14** | 2.86 | 2.57 | 2.27 | 2.85 |
| | Ni | 3.36 | 2.65 | 3.48 | 2.53 | 1.95 | **1.71** | 2.86 | 2.29 | 2.90 | 2.25 | 2.24 | 2.51 |
| | Nb | 2.28 | 2.42 | 1.68 | 3.47 | 2.58 | 2.17 | 3.06 | **1.66** | 1.78 | 2.24 | 2.05 | 2.25 |
| | Ta | 7.43 | 2.07 | 1.83 | **1.54** | 2.4 | 2.4 | 2.29 | 2.14 | 2.94 | 2.74 | 3.22 | 2.62 |
| | W | 5.75 | 3.02 | 2.49 | 2.26 | 2.78 | 2.30 | 2.28 | **1.94** | 2.34 | 2.82 | 2.96 | 2.58 |
| | Pb | 8.85 | 3.00 | 2.50 | 2.12 | 2.13 | 2.18 | 1.98 | 1.97 | **1.92** | 1.94 | 2.34 | 2.17 |
| | Bi | 5.13 | 3.40 | 2.95 | 2.12 | 1.99 | 1.82 | **1.79** | 1.97 | 2.09 | 2.10 | 2.28 | **2.12** |
| ISABEL | Cr | 5.29 | 4.45 | 3.98 | 2.49 | **2.22** | 6.30 | 4.70 | 3.85 | 2.96 | 2.63 | 2.26 | 3.58 |
| | Fe | 4.94 | 5.21 | 6.15 | 3.23 | 2.90 | 3.66 | 5.07 | 4.77 | 2.86 | 2.57 | **2.27** | 3.71 |
| | Ni | 5.45 | 3.64 | 4.14 | 2.70 | 2.21 | **2.19** | 3.62 | 4.15 | 2.90 | 2.25 | 2.24 | 3.05 |
| | Nb | 3.00 | 2.78 | 1.80 | 3.62 | 2.76 | 2.64 | 4.65 | 3.77 | **1.78** | 2.24 | 2.05 | 2.74 |
| | Ta | 7.14 | 2.24 | **1.88** | 2.29 | 2.16 | 2.54 | 2.79 | 2.56 | 2.94 | 2.74 | 3.22 | 2.75 |
| | W | 6.66 | 3.42 | 2.53 | **2.11** | 2.76 | 2.47 | 2.51 | 2.2 | 2.34 | 2.82 | 2.96 | 2.66 |
| | Pb | 15.14 | 4.09 | 2.82 | 2.00 | 2.09 | 2.16 | 2.19 | 2.07 | **1.92** | 1.94 | 2.34 | 2.28 |
| | Bi | 5.88 | 4.53 | 3.07 | 1.99 | 2.04 | **1.85** | 1.92 | 2.03 | 2.09 | 2.10 | 2.28 | **2.17** |
| CEM03.02 | Cr | 2.28 | 3.44 | 2.52 | 1.79 | 2.10 | 1.57 | **1.55** | 1.67 | 1.90 | 1.95 | 2.71 | 2.07 |
| | Fe | 3.88 | 3.98 | 2.57 | 2.12 | 2.01 | 2.34 | **1.81** | 1.83 | 1.83 | 1.96 | 2.90 | 2.24 |
| | Ni | 2.26 | 2.56 | 2.37 | **1.63** | 1.75 | 1.67 | 1.69 | 1.70 | 1.89 | 1.86 | 2.40 | 1.96 |
| | Nb | 3.36 | 2.59 | 2.08 | 2.84 | 2.26 | 1.94 | 2.09 | 1.94 | 2.67 | 2.30 | **1.92** | 2.25 |
| | Ta | 1.61 | 1.85 | 2.21 | 1.59 | **1.42** | 2.86 | 4.17 | 4.19 | 4.30 | 3.43 | 3.33 | 3.39 |
| | W | 5.52 | 3.22 | 2.88 | 2.5 | 2.32 | **1.99** | 2.13 | 2.29 | 2.87 | 2.86 | 3.31 | 2.72 |
| | Pb | 1.56 | **1.50** | 1.83 | 1.58 | 1.55 | 1.72 | 1.93 | 1.89 | 2.16 | 2.28 | 2.77 | 2.07 |
| | Bi | 4.00 | 1.65 | 1.95 | 1.76 | **1.51** | 1.81 | 1.54 | 1.99 | 2.08 | 2.10 | 2.30 | **1.97** |
| INCL4.2+ ABLA | Cr | 3.84 | 2.43 | 2.45 | 4.17 | 2.37 | 2.59 | 3.32 | 2.96 | 2.52 | **2.35** | 3.97 | 2.97 |
| | Fe | 3.17 | 3.40 | 5.10 | 3.53 | 3.61 | 4.35 | 5.31 | 6.74 | 3.18 | **3.00** | 3.91 | 4.16 |
| | Ni | 2.01 | 2.08 | **1.97** | 2.52 | 3.44 | 4.11 | 2.28 | 3.41 | 3.01 | 3.48 | 2.92 | 2.95 |
| | Nb | 2.94 | **2.02** | 2.12 | 3.99 | 3.27 | 3.26 | 3.31 | 4.90 | 4.63 | 4.40 | 3.19 | 3.77 |
| | Ta | 5.83 | 3.09 | 2.31 | 2.17 | **2.02** | 2.9 | 3.84 | 3.35 | 3.65 | 3.68 | 3.94 | 3.41 |
| | W | 6.99 | 3.8 | 2.78 | 2.57 | **2.36** | 2.93 | 3.46 | 3.06 | 4.09 | 3.78 | 3.8 | 3.47 |
| | Pb | 3.57 | 2.74 | 1.88 | **1.60** | 1.67 | 2.35 | 2.70 | 2.18 | 2.17 | 2.37 | 2.29 | **2.27** |
| | Bi | 4.99 | 4.50 | 1.83 | 1.86 | **1.64** | 2.01 | 2.52 | 1.95 | 2.33 | 2.32 | 2.37 | 2.32 |

| Codes | Tar-gets | Proton energy (GeV) | | | | | | | | | | | |
|---|---|---|---|---|---|---|---|---|---|---|---|---|---|
| | | 0.04 | 0.07 | 0.1 | 0.15 | 0.25 | 0.4 | 0.6 | 0.8 | 1.2 | 1.6 | 2.6 | Mean |
| PHITS | Cr | 5.36 | 3.32 | 4.83 | 3.69 | 3.81 | 3.03 | 2.36 | 2.19 | **2.05** | 2.16 | 2.15 | 2.89 |
| | Fe | 2.84 | 5.45 | 4.99 | 4.21 | 3.22 | 3.26 | 2.46 | 2.68 | 2.24 | **2.19** | 2.76 | 3.12 |
| | Ni | 3.68 | 2.95 | 4.11 | 3.16 | 2.45 | 2.05 | 2.12 | 1.96 | **1.67** | 1.74 | 1.76 | 2.35 |
| | Nb | 3.38 | 3.12 | 2.76 | 4.35 | 3.21 | 2.29 | 2.37 | 1.74 | 2.61 | 2.01 | **1.66** | 2.39 |
| | Ta | 2.44 | 1.91 | **1.83** | 2.49 | 3.25 | 2.27 | 2.37 | 2.08 | 2.64 | 2.51 | 2.81 | 2.51 |
| | W | 8.41 | 4.3 | 3.19 | 2.61 | 3.52 | 2.23 | 2.58 | **2.20** | 2.84 | 2.88 | 2.9 | 2.89 |
| | Pb | 2.29 | 3.05 | 2.55 | 1.97 | 2.23 | 2.28 | 2.28 | 1.91 | **1.71** | 1.77 | 1.87 | **2.02** |
| | Bi | 9.72 | 4.46 | 3.41 | 2.09 | 1.95 | 2.28 | 2.09 | 2.02 | 1.92 | **1.87** | 1.88 | 2.12 |
| INCL4.5+ ABLA07 | Cr | 2.24 | 1.75 | **1.58** | 1.61 | 1.83 | 1.58 | 1.70 | 1.85 | 1.83 | 1.76 | 2.09 | 1.80 |
| | Fe | 2.49 | 3.44 | 1.81 | 1.78 | 2.62 | 2.63 | 1.75 | **1.62** | 1.69 | 1.65 | 2.11 | 2.06 |
| | Ni | 2.66 | 1.62 | 1.59 | **1.46** | 1.90 | 1.57 | 1.56 | 1.59 | 1.54 | 1.64 | 1.85 | **1.71** |
| | Nb | 13.78 | 2.05 | 3.45 | 1.98 | **1.55** | 1.65 | 1.63 | 1.56 | 1.59 | 1.55 | 1.58 | 1.93 |
| | Ta | 3.89 | 2.09 | 2.33 | 1.81 | **1.48** | 1.59 | 1.85 | 1.75 | 1.8 | 1.73 | 2.01 | 1.83 |
| | W | 6.05 | 4.34 | 2.66 | 2.14 | **1.73** | 1.84 | 2.11 | 1.76 | 2.03 | 1.89 | 2.01 | 2.13 |
| | Pb | 2.18 | 2.20 | 1.92 | 1.66 | 1.69 | 1.56 | 1.49 | **1.47** | 1.52 | 1.57 | 1.61 | **1.60** |
| | Bi | 3.71 | 2.62 | 1.87 | 1.61 | 1.63 | 1.61 | **1.48** | 1.56 | 1.56 | 1.58 | 1.60 | 1.63 |
| CASCA-DE07 | Cr | **2.61** | 3.27 | 3.02 | 2.87 | 3.26 | 2.93 | 2.78 | 2.80 | 5.34 | 5.04 | 4.93 | 3.67 |
| | Fe | 5.92 | 4.23 | 3.33 | **2.93** | 3.57 | 3.96 | 3.35 | 3.24 | 5.47 | 4.90 | 4.54 | 3.98 |
| | Ni | 5.19 | 3.45 | 3.47 | 3.20 | 2.75 | 2.77 | 2.65 | **2.60** | 4.37 | 4.13 | 3.83 | 3.44 |
| | Nb | 9.97 | 4.73 | 4.33 | 3.37 | 2.65 | 2.61 | 2.87 | **2.37** | 2.73 | 3.69 | 3.01 | 3.19 |
| | Ta | 9.43 | **1.70** | 2.79 | 1.75 | 2.94 | 3.16 | 2.65 | 2.42 | 3.11 | 3.17 | 2.96 | 2.89 |
| | W | 6.26 | 3.3 | 3.24 | 2.8 | 2.97 | 2.45 | **2.42** | 2.48 | 3.25 | 3.2 | 3.34 | **3.02** |
| | Pb | 2.87 | 6.37 | 4.61 | 4.38 | 4.69 | 3.88 | 2.82 | 2.86 | **2.54** | 2.99 | 3.70 | 3.40 |
| | Bi | **1.63** | 25.53 | 6.14 | 4.97 | 4.48 | 4.84 | 3.35 | 3.39 | 3.09 | 3.02 | 3.38 | 3.84 |

The estimated deviation factors can be considered as an indicator of the predictive power of HETC. A comparison of such deviation factors in the whole incident proton energy range is shown in Figs. 3-10 for each of the targets considered here [60], respectively.

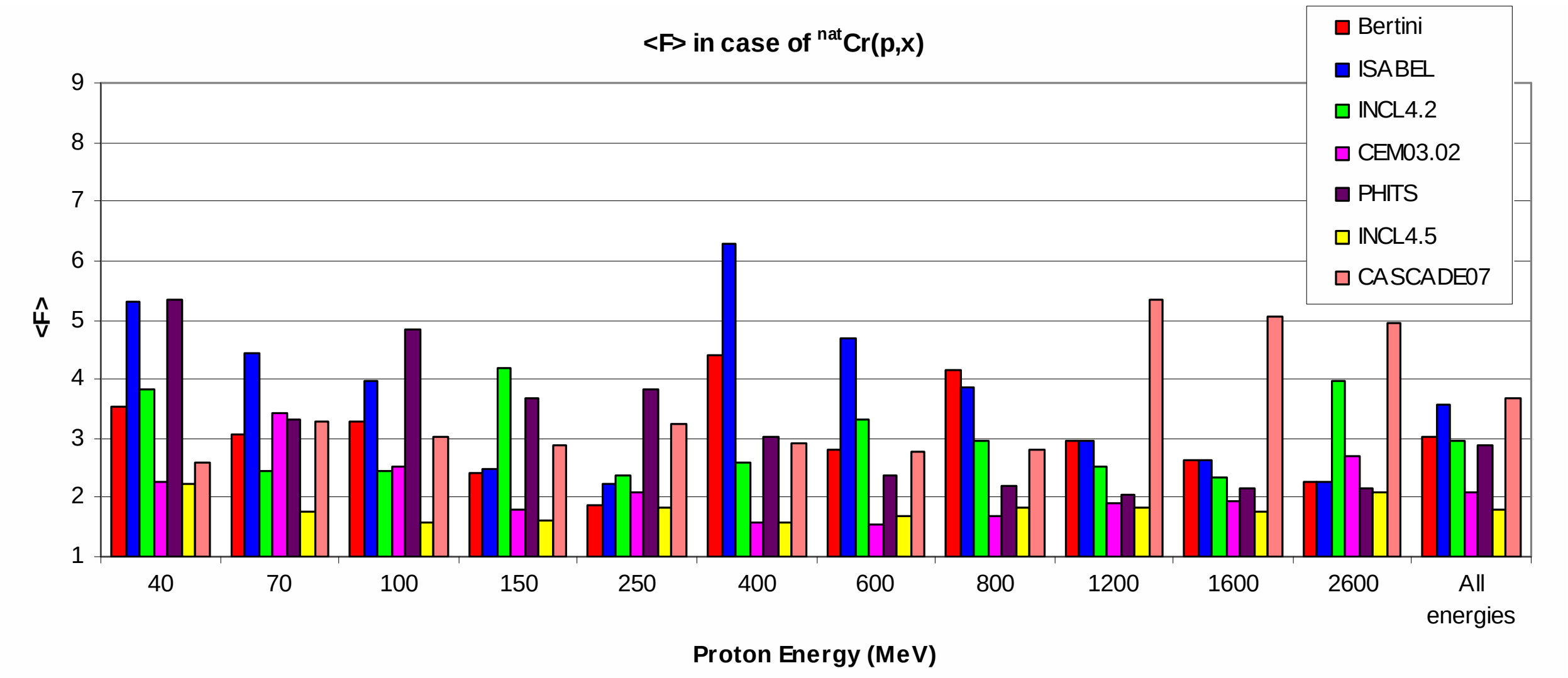

Fig. 3. Predictive accuracy of HETC for the $^{nat}$Cr target.

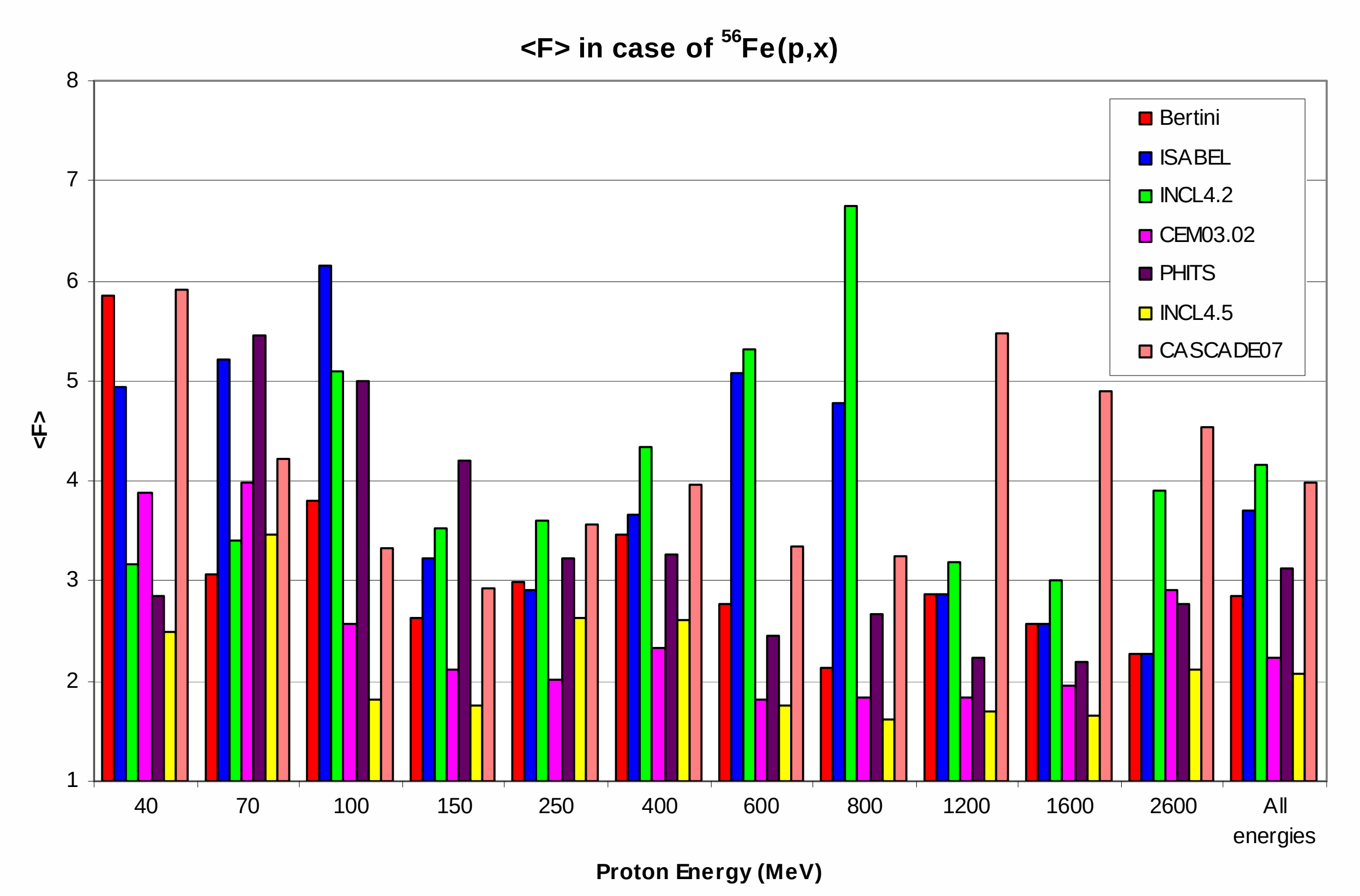

Fig. 4. The same as in Fig. 3, but $^{56}Fe$.

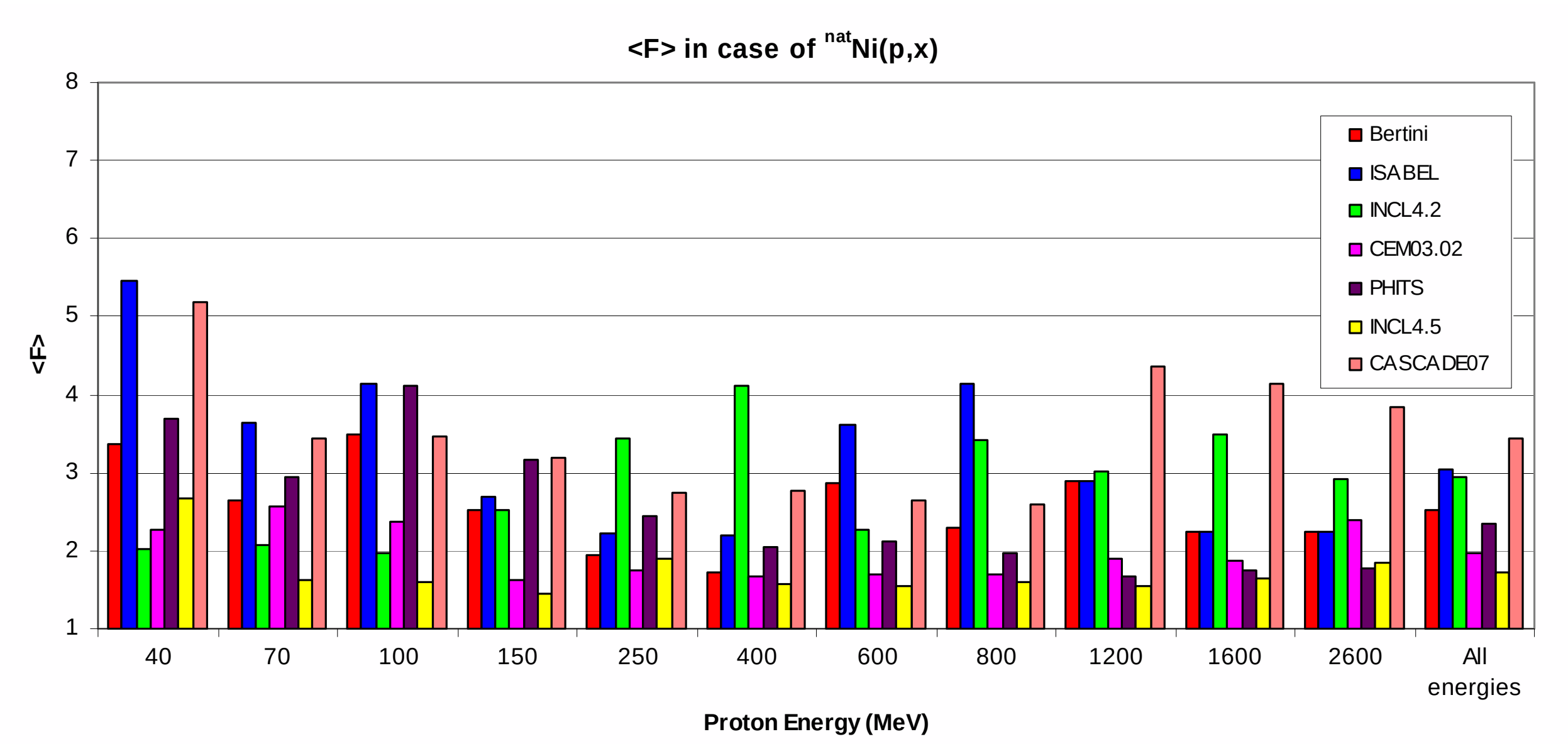

Fig. 5. The same as in Fig. 3, but for $^{nat}Ni$.

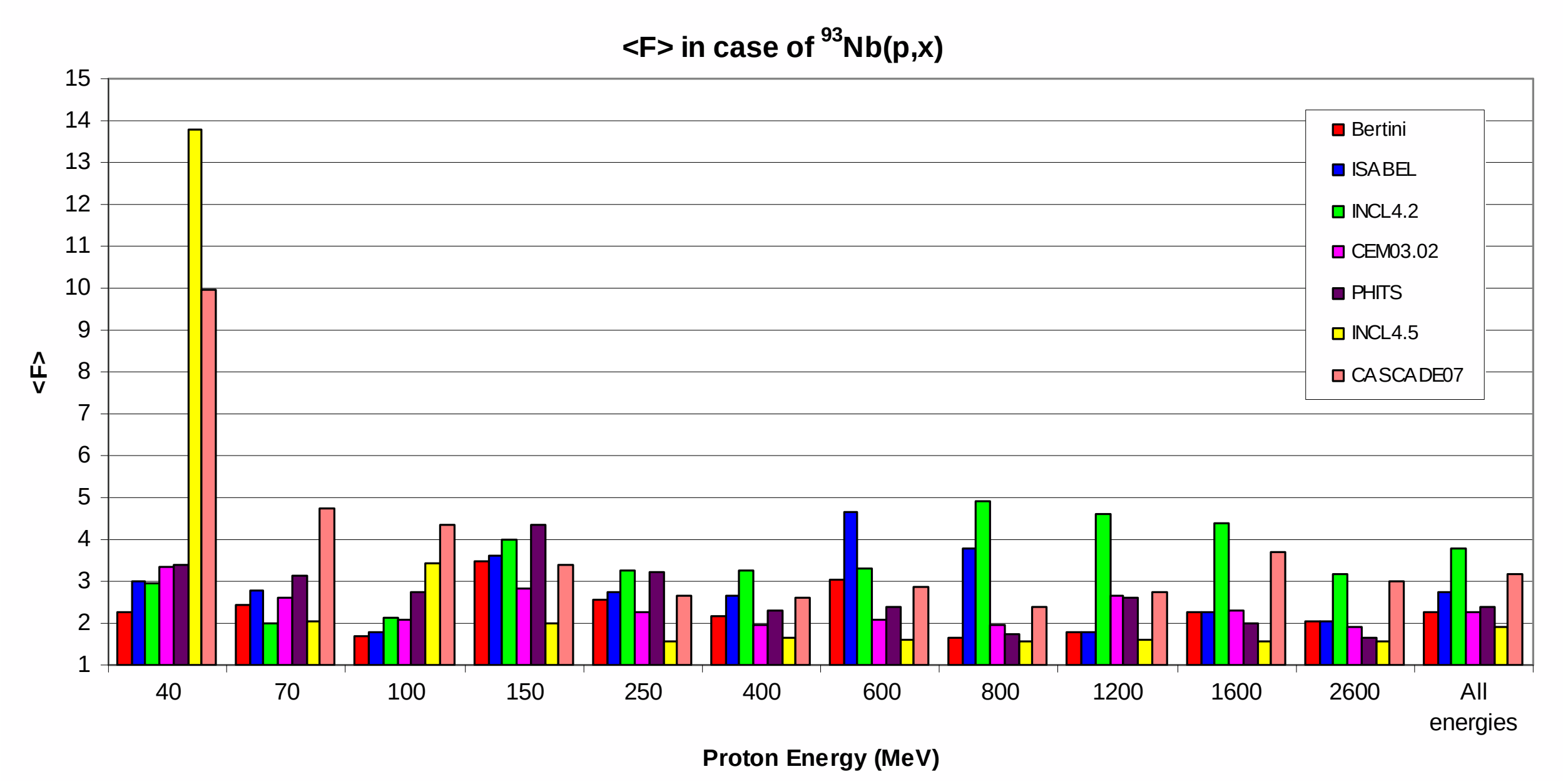

Fig. 6. The same as in Fig. 3, but for $^{93}$Nb.

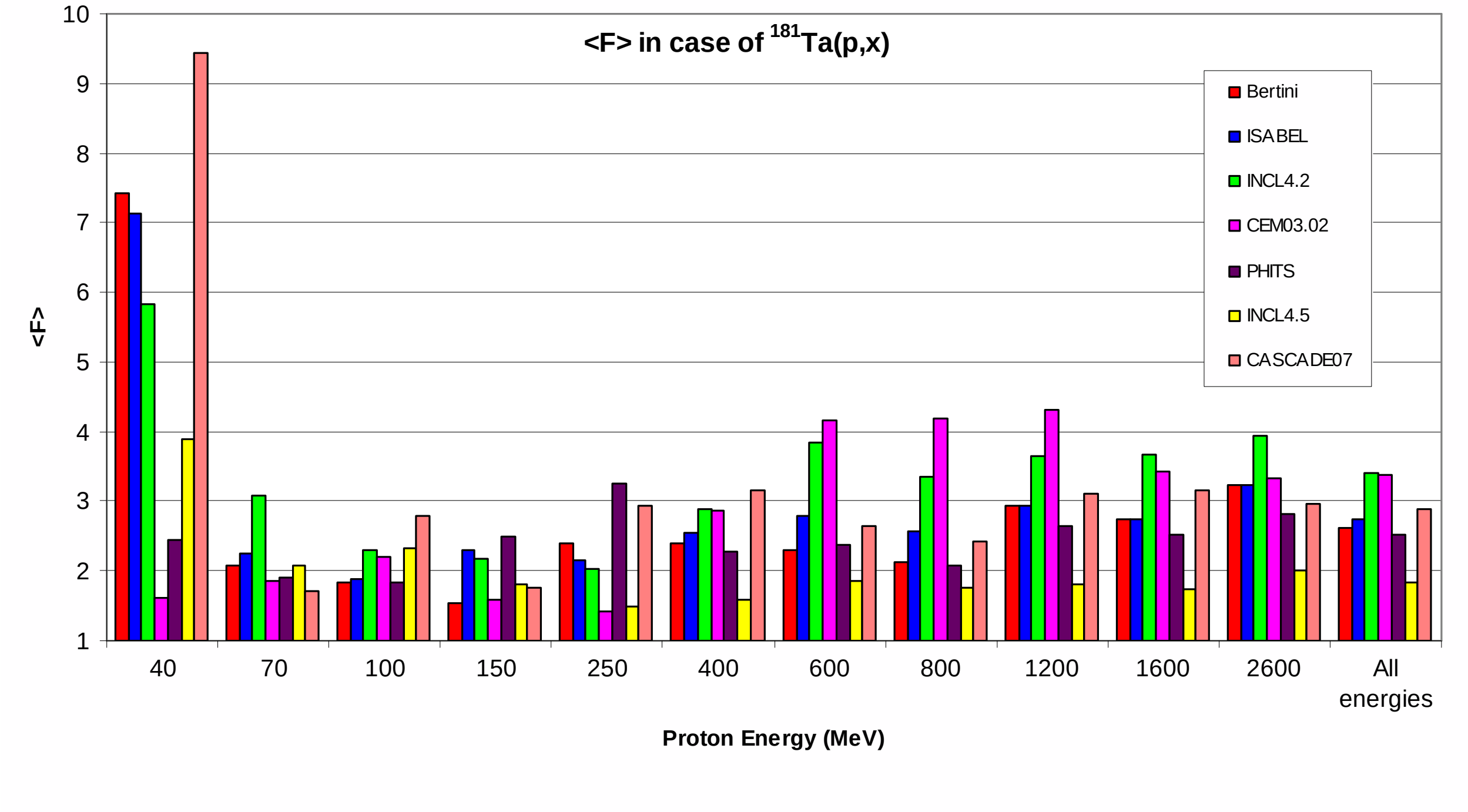

Fig. 7. The same as in Fig. 3, but for $^{181}$Ta.

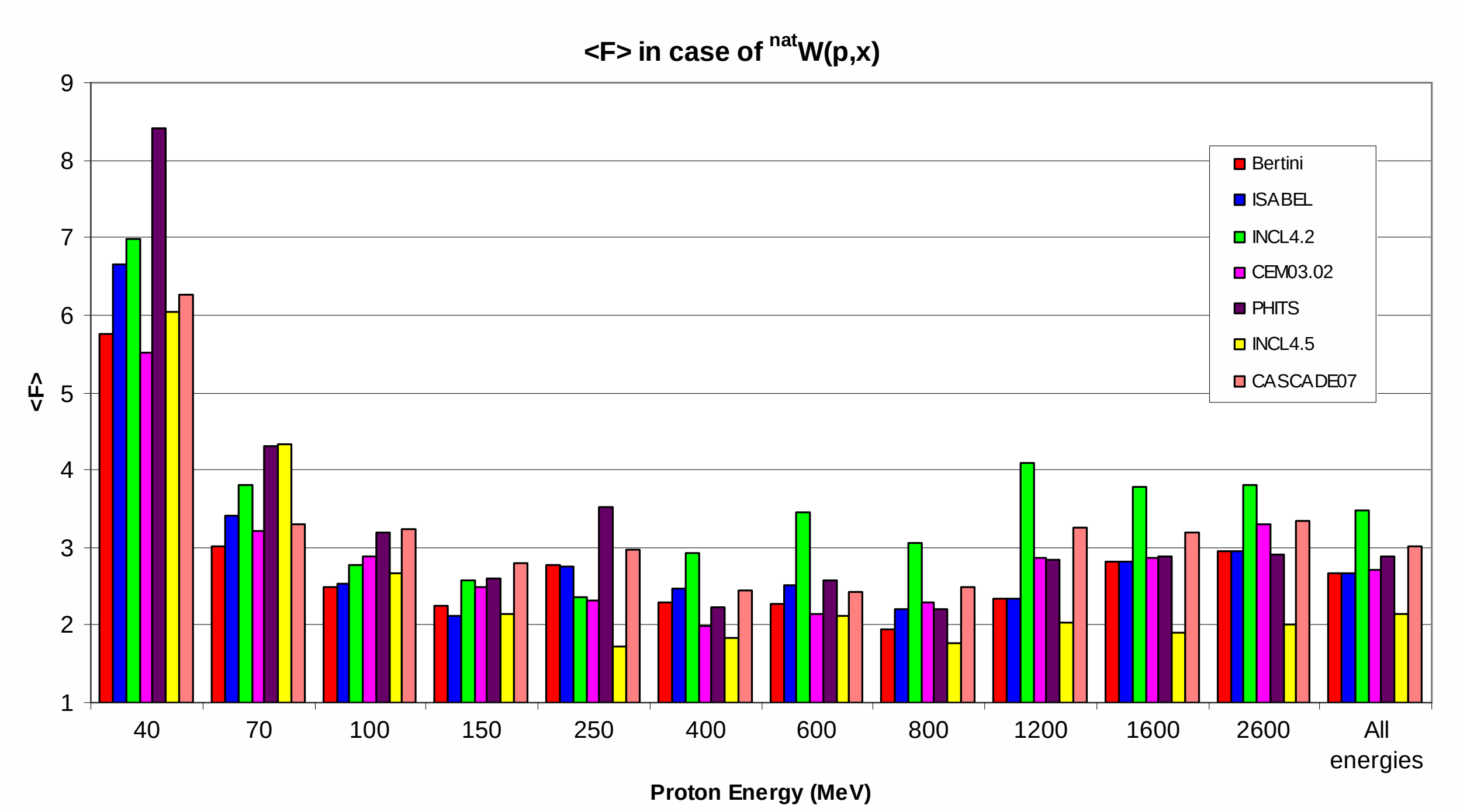

Fig. 8. The same as in Fig. 3, but for $^{nat}$W.

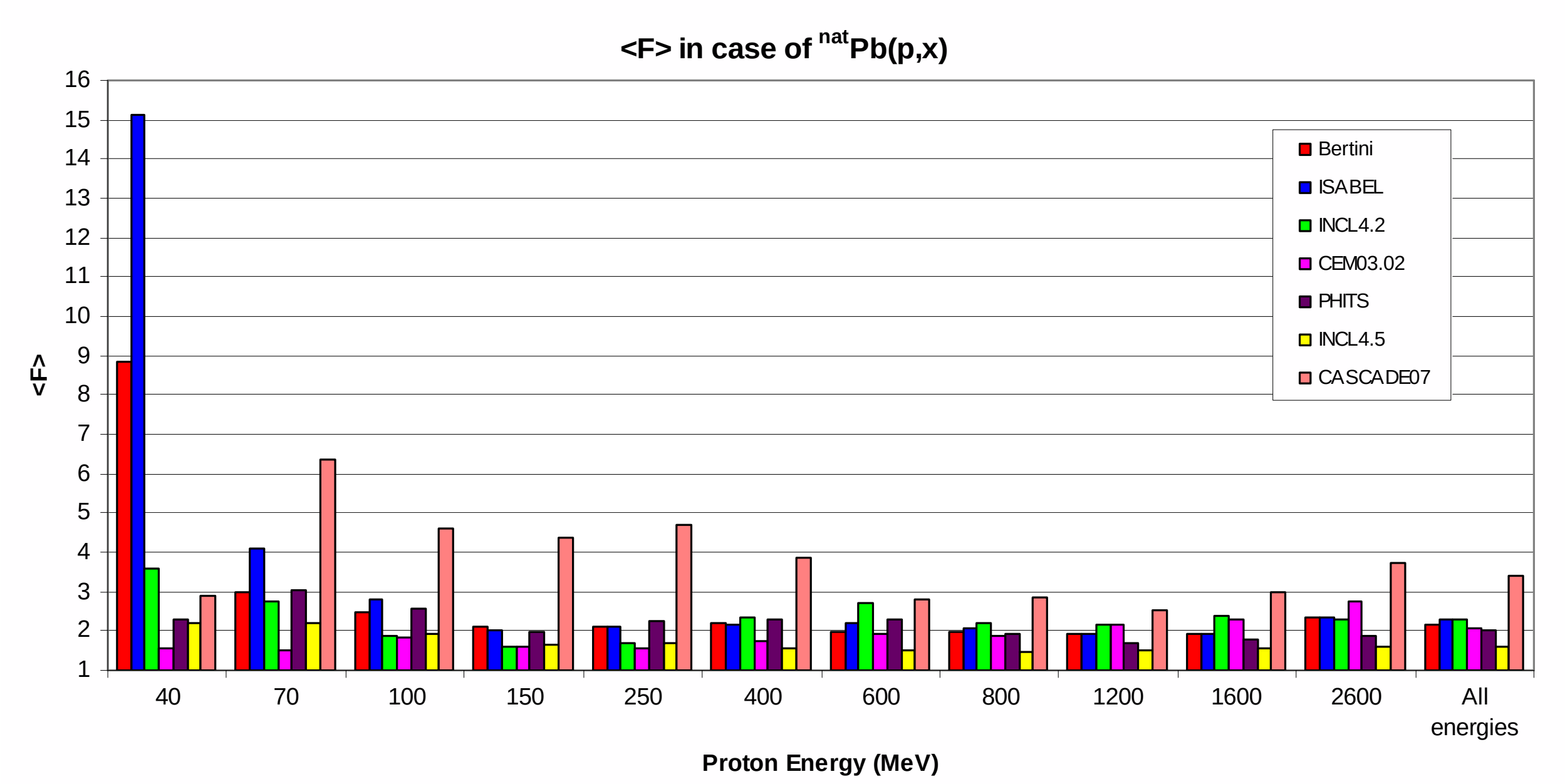

Fig. 9. The same as in Fig. 3, but for $^{nat}$Pb.

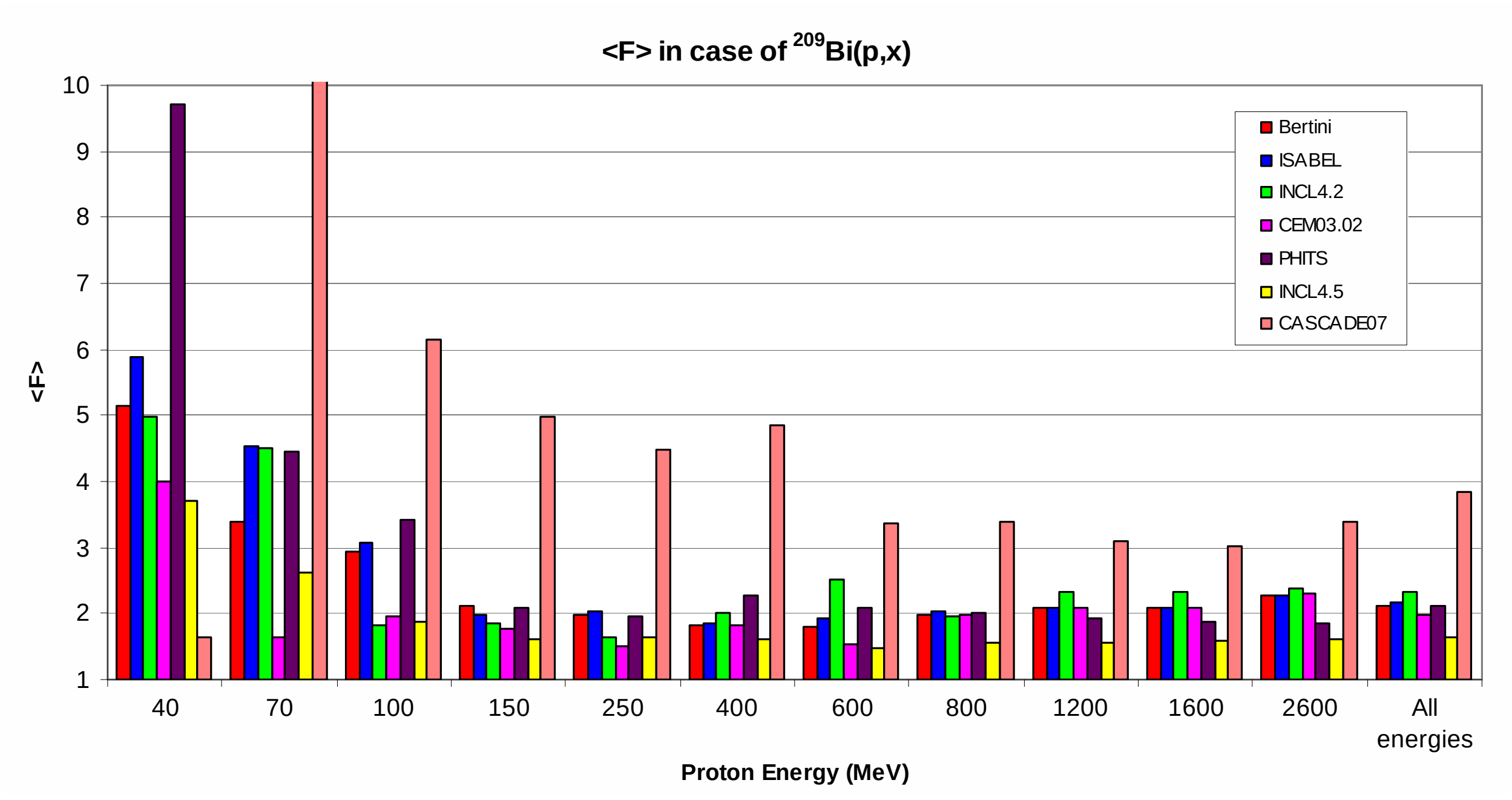

Fig. 10. The same as in Fig. 3, but for $^{209}$Bi.

To estimate the predictive accuracy of the codes for the whole mass region, including the targets not covered by our measurements, and for all incident proton energies, including the intermediate ones, the calculated deviation factors are presented in Figs. 11-17 in the form of a two-dimensional diagram for each code [60], respectively.

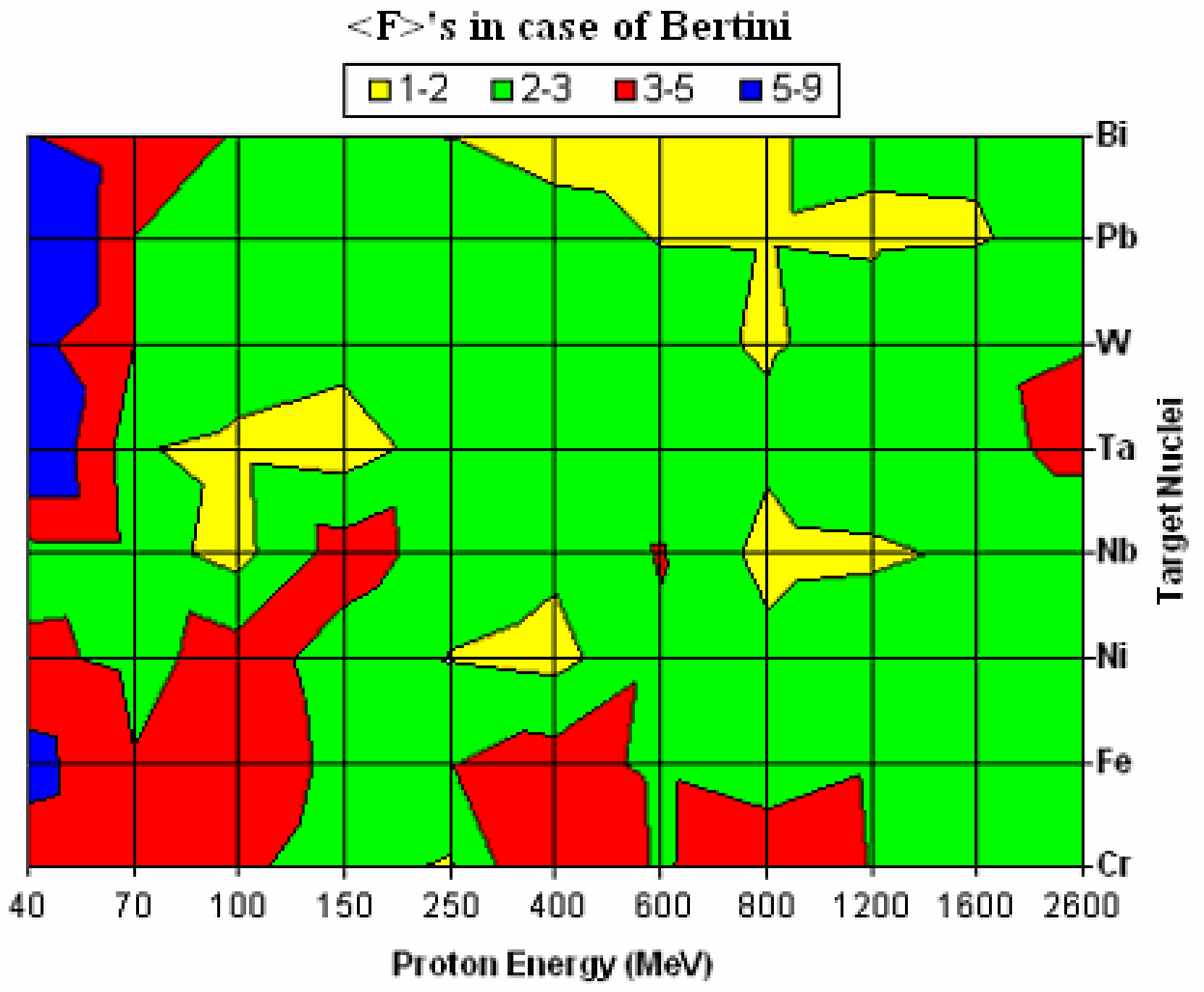

Fig. 11. Predictive accuracy of the Bertini code for various proton energies and target masses.

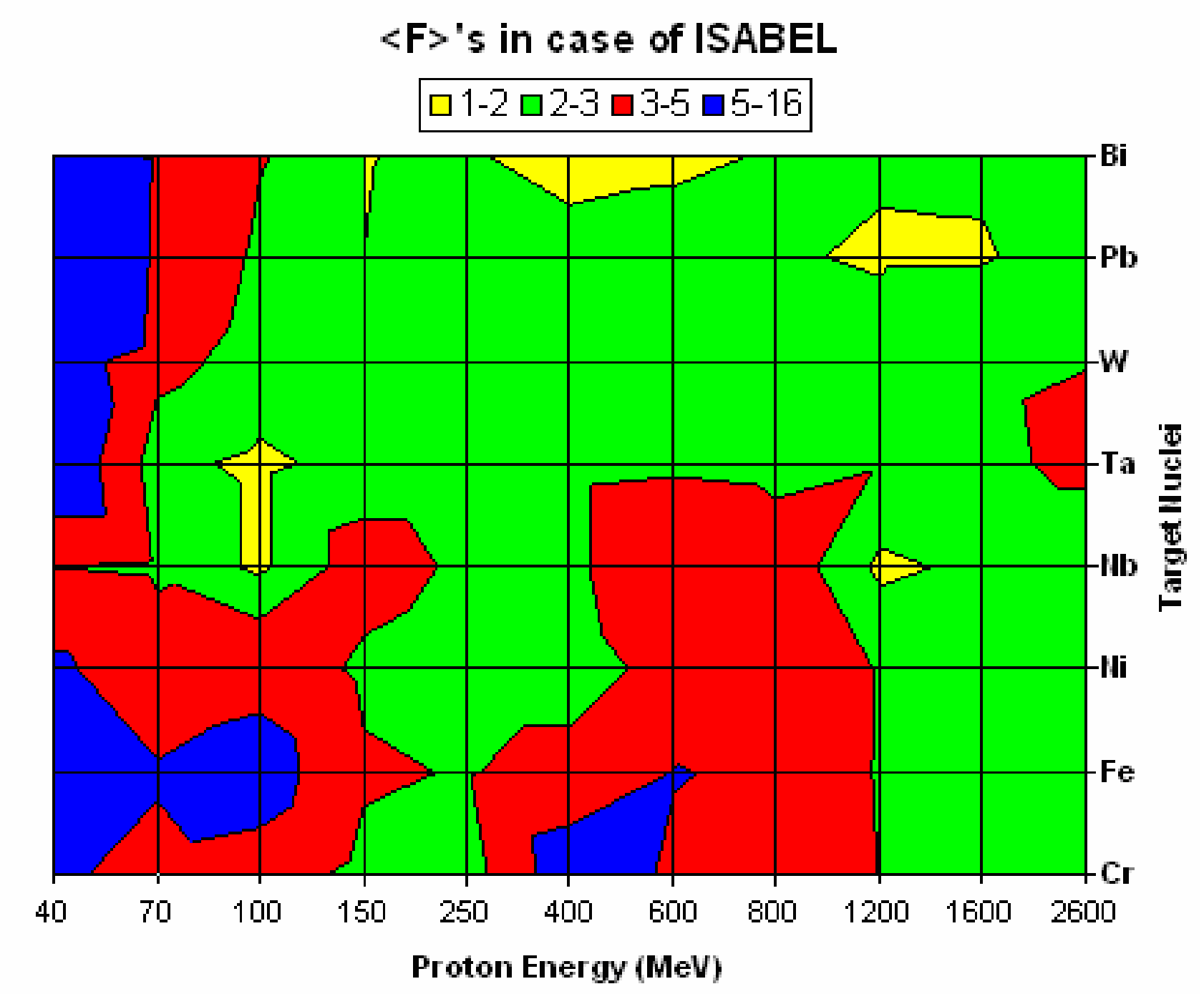

Fig. 12. The same as in Fig. 11, but for the ISABEL code.

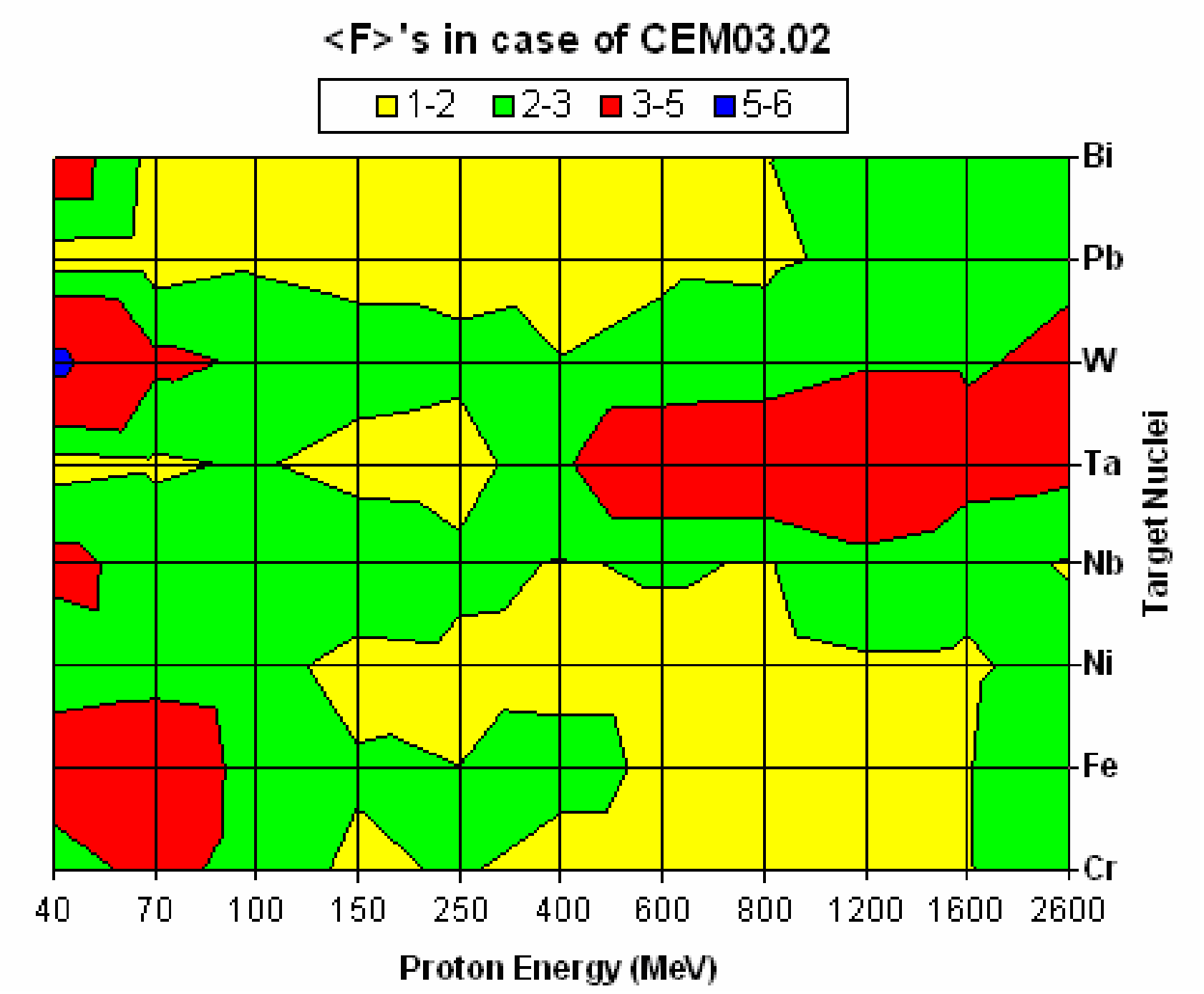

Fig. 13. The same as in Fig. 11, but for the CEM03.02 code.

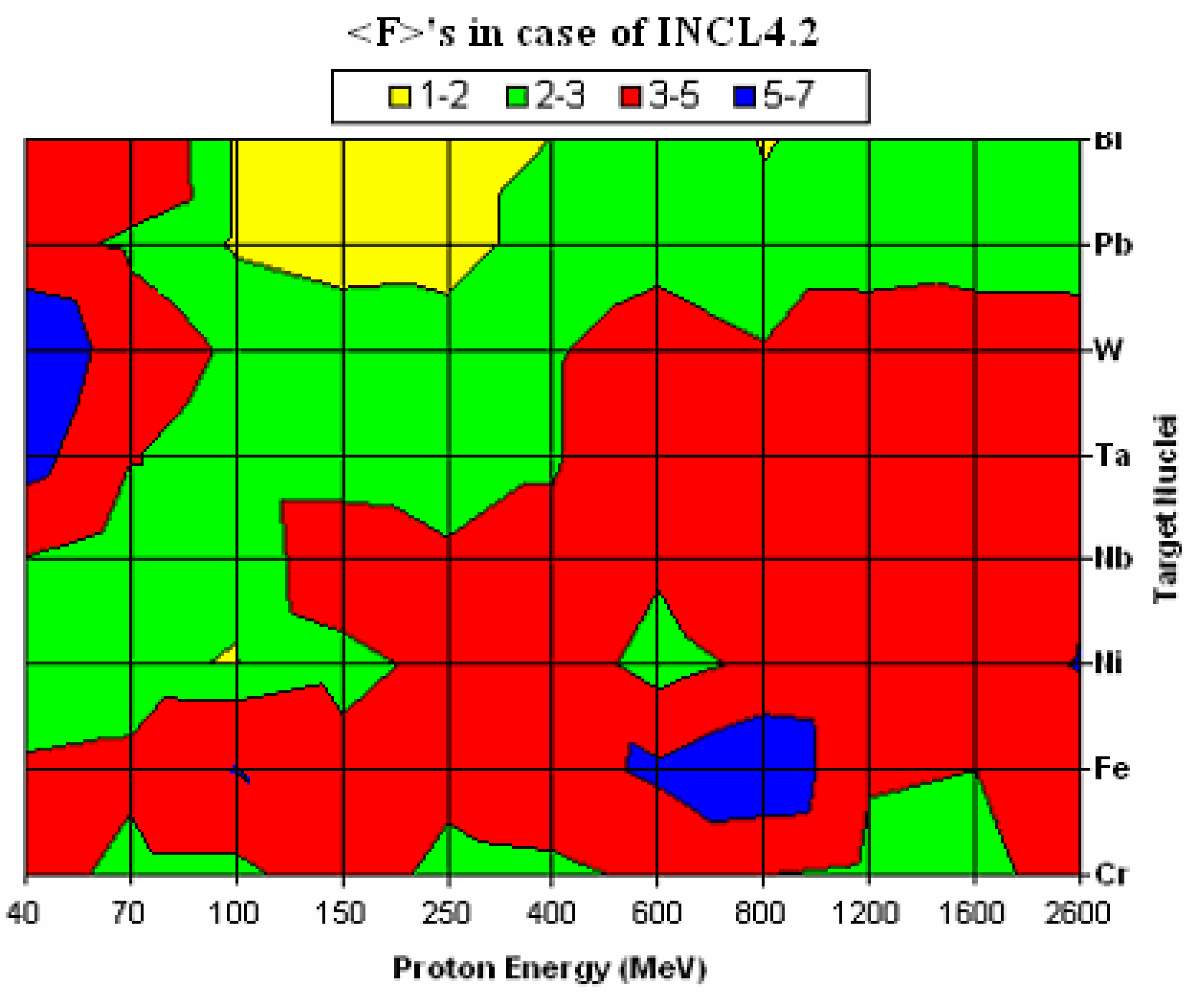

Fig. 14. The same as in Fig. 11, but for the INCL4.2+ABLA code.

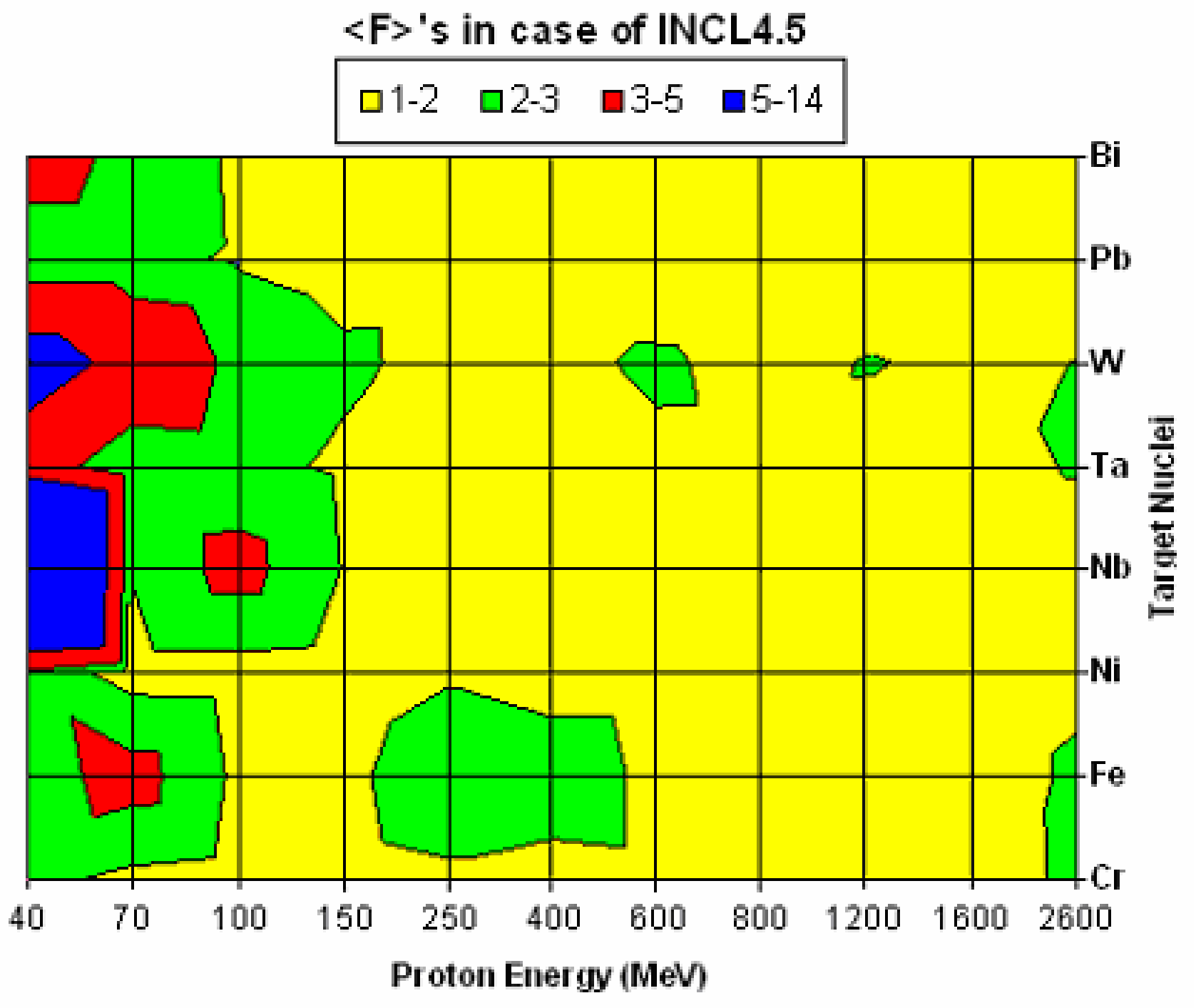

Fig. 15. The same as in Fig. 11, but for the INCL4.5+ABLA07 code.

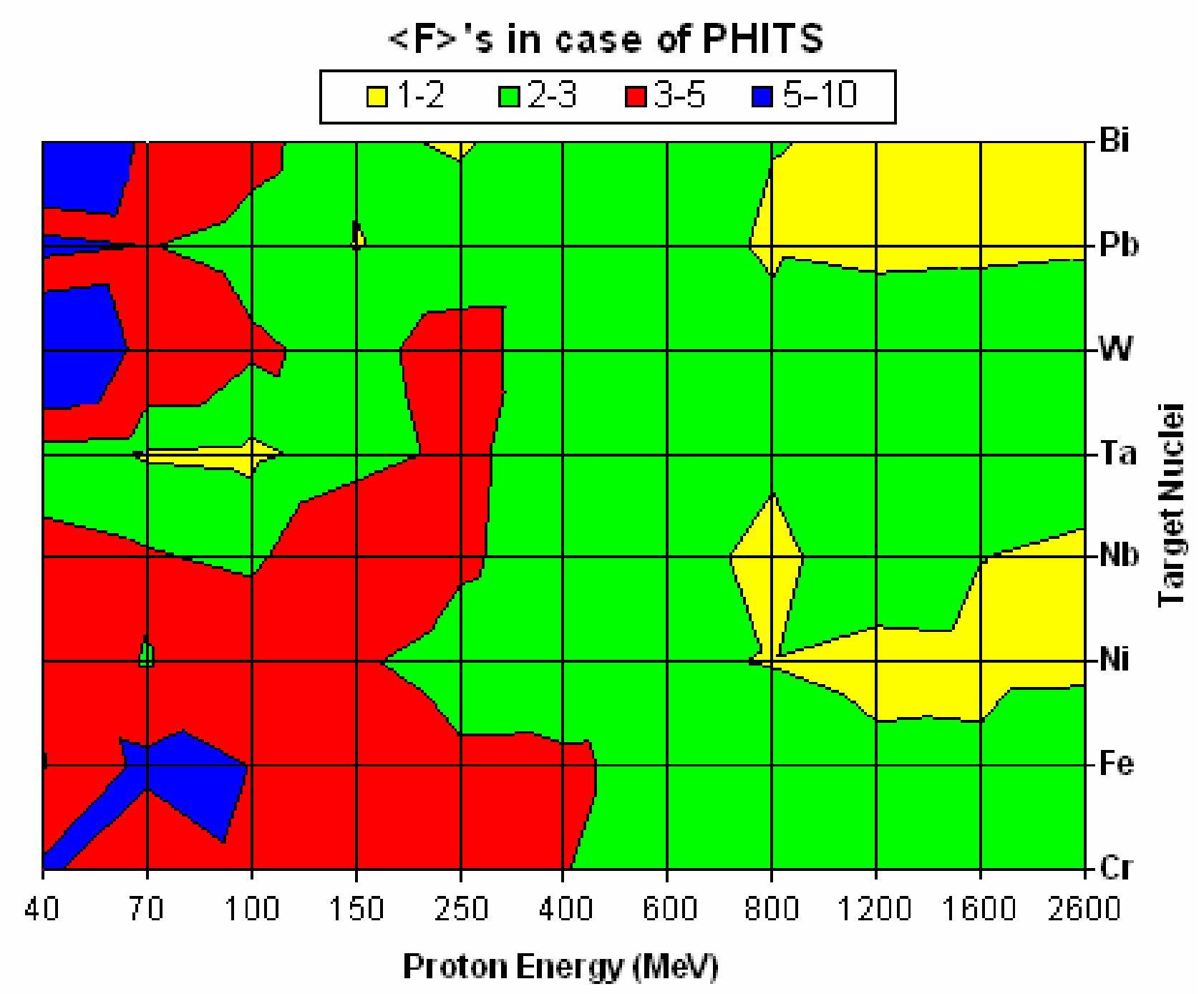

Fig. 16. The same as in Fig. 11, but for the PHITS code.

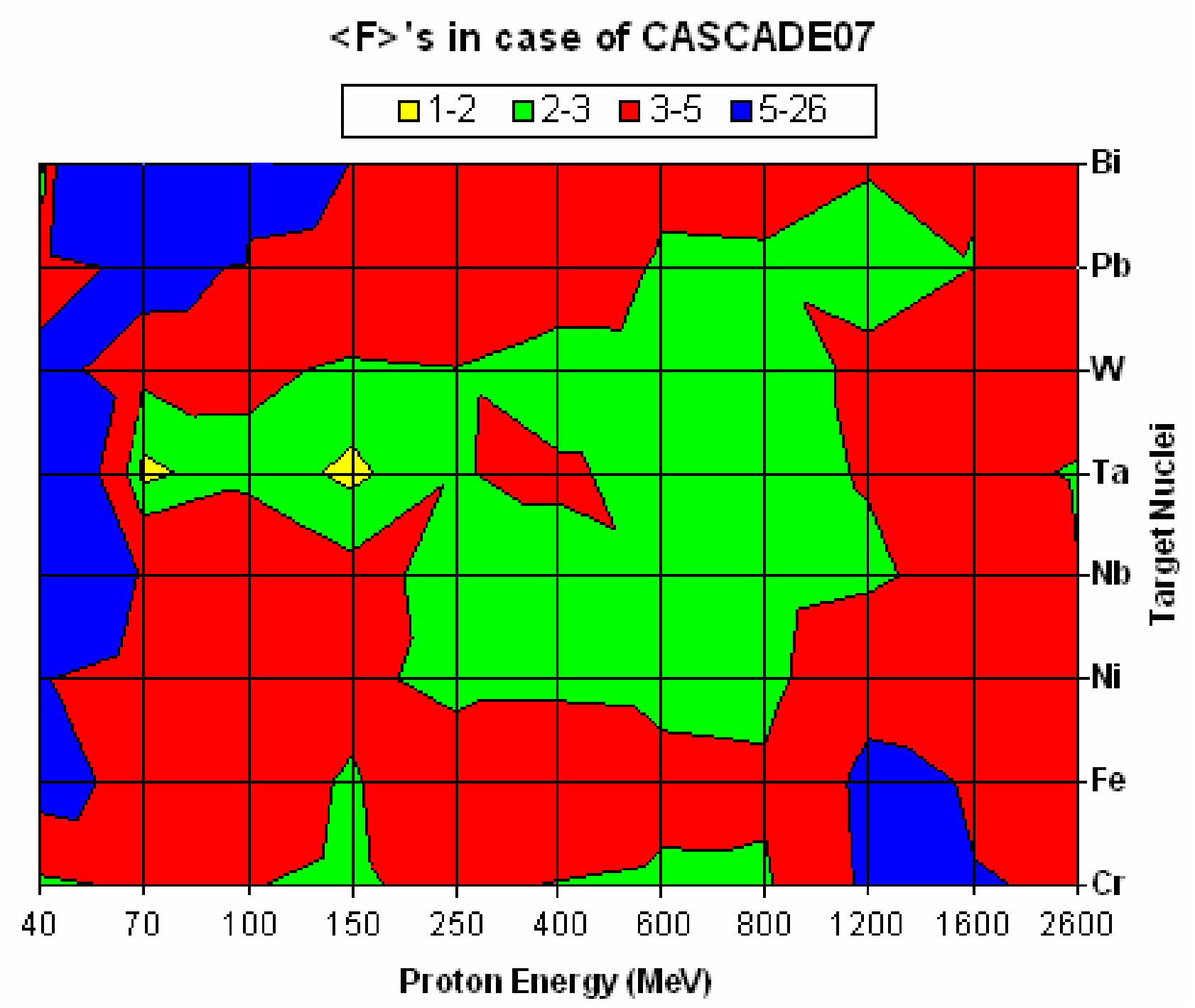

Fig. 17. The same as in Fig. 11, but for CASCADE07 code.

The results presented above allow us to estimate the regions of target mass numbers and proton incident energies where the predictive accuracy of every code is not good enough:

Bertini: ~ 2600 MeV (W, Ta), 40 -100 MeV (Nb, Fe, Cr), 40-70 MeV (Bi, Pb, W, Ta, Nb);

ISABEL: 2600 MeV (W, Ta), 250-1200 MeV (Ta, Nb, Ni, Fe, Cr ), 70-150 MeV (Nb, Ni, Fe, Cr), 40-70MeV (W, Ta, Nb, Ni, Fe, Cr), 40-100 MeV (Bi, Pb);
CEM03.02: 500 -2600 (W, Ta), 40 – 100(Fe, Cr), 40(Bi, Nb), 40-70(W);
INCL4.2+ABLA: 400-2600 (W, Ta); 150-400 (Nb, Ni, Fe, Cr); 40 150 (Ni, Fe, Cr); 40-100 (Bi, Pb, W, Ta, Nb);
INCL4.5+ABLA07: 40-70(W, Ta, Nb, Ni), 40 (Bi);
PHITS: 40-400(Fe, Cr), 40-150 (Nb, Ni), 100-250 (W, Ta, Nb), 40-100 (Bi, Pb, W);
CASCADE07: 1600-2600 (Bi, Pb, W, Ta, Nb, Ni, Fe, Cr), 1200-1600 (W, Ta, Nb, Ni, Fe, Cr), 250-400 (Fe, Cr), 40-250 (Ni, Fe, Cr), 40-150 (Nb, Ni, Fe, Cr), 70-150 (Bi, Pb, W), 40-70 (Bi, Pb, W, Ta, Nb, Ni, Fe, Cr)
The required accuracy of nuclear data for a majority of practical applications is about 30-50%. Such accuracy corresponds to deviation factor values of 1.3-1.5. In accordance with Table V, accuracies close to the required ones are achieved with the CEM03.02 and NCL4.5+ABLA07 codes only for lead and bismuth targets, experimental data of which have been widely used to optimize these codes. In applications of these codes to other mass regions, their predictive powers are still below the required one. The predictive powers of other codes tested here are even worse.

## VI. CONCLUSIONS

The estimated predictive accuracy of the most popular HETC tested here indicates that all codes need further improvements. Nowadays, the CEM03.02 (as developed during 2004-2006) and NCL4.5+ABLA07 (as developed during 2008-2009) codes can be considered as the most accurate, due to many recent improvements of them in the processes of the GSI data analyses. The whole set of experimental data obtained by the activation methods for many targets and a wide range of excitation energies opens additional opportunities for the tests and improvements of the models included in the current HETC systems. Last, let us mention that codes developed for practical applications must not only describe reasonably well arbitrary nuclear reactions without any free parameters, but also not require too much computing time. As calculations for this study, as well as for the recent benchmark of the spallation models [3], were performed by different persons at different computers, the question about the computing time required by various codes remains open.

## ACKNOWLEDGEMENTS

The authors are indebted to Prof. Rolf Michel (ZSR, Hamburg) for his recommendations on monitor reactions and many helpful general remarks and to Prof. Cornelis H. M. Broeders (KFK,

Karlsruhe) for informative discussions on results of our ISTC Projects. We gratefully acknowledge Drs. Konstantin K. Gudima (Institute of Applied Physics, Academy of Science of Moldova, Kishinev, Moldova) and Arnold J. Sierk (Los Alamos National Laboratory (LANL), USA), both co-authors of the CEM03.03 and CEM03.02 codes, for useful discussions and for performing calculations for the recent benchmark of spallation models [3] with the CEM03.03 and CEM03.02 codes, respectively.

The present work has been carried out under the ISTC Projects 2002 and 3266 and was supported permanently by the ITEP and Rosatom authorities. Part of the work performed at LANL was carried out under the auspices of the National Nuclear Security Administration of the US Department of Energy at Los Alamos National Laboratory under Contract No. DE-AC52-06NA25396.

## Appendix

To extend the amount of data important for a wide comparison of the results obtained by the direct and inverse kinematics methods, additional irradiation of a $^{208}$Pb (97.2%) target with a 500-MeV proton beam has been carried out in ITEP. The obtained cross sections for the production of residual radioactive nuclides are presented in Table VI. A comparison of these data with the cross sections measured at GSI in the inverse kinematics approach [27, 28] and converted here to cumulative yields as measured at ITEP is shown in Fig. 18.

TABLE VI. Nuclides production cross sections measured at ITEP with the activation method for a target of $^{208}$Pb irradiated by 500 MeV protons.

| Nuclide | Type | $T_{1/2}$ | $\sigma \pm \Delta\sigma$ | Nuclide | Type | $T_{1/2}$ | $\sigma \pm \Delta\sigma$ |
|---|---|---|---|---|---|---|---|
| $^{207}$Bi | i | 31.55y | 4.31 ± 0.64 | $^{183}$Ir | c | 57m | 20.7±5.2 |
| $^{206}$Bi | i | 6.243d | 6.16±0.46 | $^{185}$Os | c | 93.6d | 33.8±2.6 |
| $^{205}$Bi | i | 15.31d | 8.74±0.72 | $^{183m}$Os | c* | 9.9h | 14.5±1.1 |
| $^{204}$Bi | i | 11.22h | 7.88±0.58 | $^{182}$Os | c | 22.10h | 26.7±2.0 |
| $^{203}$Bi | i(m+g) | 11.76h | 8.28±0.68 | $^{181}$Os | c | 105m | 7.7±1.1 |
| $^{202}$Bi | i | 1.72h | 6.87±0.59 | $^{180}$Os | c | 21.5m | 12.3±1.1 |
| $^{204m}$Pb | i(m) | 67.2m | 13.9±1.1 | $^{183}$Re | c | 70.0d | 26.5±2.2 |
| $^{204m}$Pb | c | 67.2m | 15.1±1.1 | $^{182m}$Re | c | 12.7h | 25.3±1.9 |
| $^{203}$Pb | c* | 51.873h | 45.5±3.8 | $^{181}$Re | c* | 19.9h | 17.3±2.4 |
| $^{202m}$Pb | i(m) | 3.53h | 15.9±1.7 | $^{179}$Re | c* | 19.5m | 15.7±1.7 |
| $^{201}$Pb | c* | 9.33h | 36.8±3.5 | $^{178}$Re | c* | 13.2m | 11.7±1.7 |
| $^{200}$Pb | c | 21.5h | 32.5±2.6 | $^{178}$W | c | 21.6d | 9.8±1.3 |
| $^{199}$Pb | c* | 90m | 48.6±5.6 | $^{177}$W | c | 135m | 6.45±0.84 |
| $^{198}$Pb | c | 2.4h | 43.3±7.0 | $^{177}$Ta | c* | 56.56h | 9.0±2.1 |
| $^{197m}$Pb | c* | 43m | 30.9±2.9 | $^{176}$Ta | c | 8.09h | 8.25±0.90 |
| $^{196}$Pb | c* | 37m | 19.3±1.9 | $^{175}$Hf | c | 70d | 5.26±0.43 |
| $^{195m}$Pb | i(m) | 15.0m | 18.3±2.5 | $^{172}$Hf | c | 1.87y | 2.03±0.18 |
| $^{202}$Tl | c | 12.23d | 21.6±1.6 | $^{173}$Lu | c | 1.37y | 2.89±0.26 |
| $^{201}$Tl | c* | 72.912h | 71.9±5.3 | $^{171}$Lu | c* | 8.24d | 2.34±0.56 |
| $^{200}$Tl | i | 26.1h | 28.7±2.4 | $^{170}$Lu | c | 2.012d | 1.21±0.15 |
| $^{200}$Tl | c | 26.1h | 60.4±4.4 | $^{169}$Lu | c | 34.06h | 1.08±0.11 |
| $^{199}$Tl | c* | 7.42h | 62.7±6.5 | $^{169}$Yb | c* | 32.026d | 0.846±0.098 |
| $^{198m}$Tl | i(m) | 1.87h | 28.3±3.8 | $^{139}$Ce | c | 137.640d | 0.204±0.017 |
| $^{198}$Tl | c | 5.3h | 48.2±4.8 | $^{127}$Xe | c | 36.4d | 0.485±0.039 |
| $^{197}$Tl | c* | 2.84h | 86±28 | $^{123m}$Te | i(m) | 119.7d | 0.379±0.029 |
| $^{196m}$Tl | i(m) | 1.41h | 47.2±8.8 | $^{121m}$Te | i(m) | 154d | 0.391±0.034 |
| $^{195}$Tl | c* | 1.16h | 39.9±5.4 | $^{121}$Te | c | 19.16d | 0.624±0.063 |
| $^{203}$Hg | c | 46.612d | 3.73±0.28 | $^{119m}$Te | i(m) | 4.70d | 0.216±0.018 |
| $^{197m}$Hg | i(m) | 23.8h | 12.70±1.00 | $^{120m}$Sb | i(m) | 5.76d | 0.483±0.058 |
| $^{195m}$Hg | i(m) | 41.6h | 16.6±1.5 | $^{117m}$Sn | i(m) | 13.60d | 0.750±0.057 |
| $^{195}$Hg | i | 9.9h | 57.4±8.6 | $^{114m}$In | i(m) | 49.51d | 1.110±0.100 |
| $^{195}$Hg | c | 9.9h | 67.6±9.7 | $^{110m}$Ag | i(m) | 249.76d | 1.220±0.090 |
| $^{193m}$Hg | i(m) | 11.8h | 16.3±5.7 | $^{106m}$Ag | i(m) | 8.28d | 0.352±0.033 |
| $^{193}$Hg | c | 3.80h | 24.5±3.7 | $^{105}$Rh | c | 35.36h | 4.78±0.39 |
| $^{192}$Hg | c | 4.85h | 59.9±7.9 | $^{102}$Rh | i | 207d | 0.476±0.072 |
| $^{191}$Hg | c | 49m | 37±20 | $^{101m}$Rh | c | 4.34d | 0.616±0.063 |
| $^{190}$Hg | c* | 20.0m | 32.5±3.6 | $^{101}$Rh | c | 3.3y | 0.335±0.073 |
| $^{198m}$Au | i(m) | 2.27d | 0.711±0.068 | $^{103}$Ru | c | 39.26d | 4.58±0.34 |
| $^{198}$Au | i(m+g) | 2.69517d | 2.01±0.15 | $^{96}$Tc | i(m+g) | 4.28d | 0.528±0.042 |
| $^{198}$Au | i | 2.69517d | 1.280±0.100 | $^{99}$Mo | c | 65.94h | 3.66±0.58 |
| $^{196}$Au | i(m1+m2+g) | 6.183d | 3.82±0.28 | $^{96}$Nb | i | 23.35h | 2.42±0.29 |
| $^{195}$Au | c | 186.098d | 83.7±7.6 | $^{95}$Nb | i(m+g) | 34.975d | 2.98±0.22 |
| $^{194}$Au | i(m1+m2+g) | 38.02h | 6.36±0.62 | $^{95}$Nb | c | 34.975d | 5.92±0.43 |
| $^{193}$Au | c | 17.65h | 69.6±7.2 | $^{92m}$Nb | i(m) | 10.15d | 0.273±0.021 |
| $^{192}$Au | i(m+g) | 4.94h | 8.6±1.3 | $^{97}$Zr | c | 16.744h | 0.786±0.062 |
| $^{191}$Au | c* | 3.18h | 61.2±4.4 | $^{95}$Zr | c | 64.02d | 2.96±0.21 |
| $^{190}$Au | i | 42.8m | 15.1±2.5 | $^{89}$Zr | c | 78.41h | 0.949±0.070 |
| $^{190}$Au | c | 42.8m | 48.1±6.1 | $^{88}$Zr | c | 83.4d | 0.354±0.025 |
| $^{191}$Pt | c | 2.802d | 54.6±5.1 | $^{90m}$Y | i(m) | 3.19h | 2.93±0.23 |
| $^{189}$Pt | c | 10.87h | 49.7±3.8 | $^{88}$Y | i | 106.65d | 1.81±0.19 |
| $^{188}$Pt | c | 10.2d | 46.2±3.8 | $^{88}$Y | c | 106.65d | 2.15±0.16 |
| $^{187}$Pt | c | 2.35h | 27.8±5.0 | $^{87}$Y | c* | 79.8h | 1.14±0.14 |
| $^{186}$Pt | c | 2.08h | 24.8±4.6 | $^{85}$Sr | c | 64.84d | 1.35±0.11 |
| $^{184}$Pt | c | 17.3m | 18.2±3.7 | $^{86}$Rb | i(m+g) | 18.631d | 3.76±0.30 |
| $^{192}$Ir | i(m1+g) | 73.827d | 0.098±0.010 | $^{83}$Rb | c | 86.2d | 1.49±0.13 |

| Nuclide | Type | $T_{1/2}$ | σ ± Δσ | Nuclide | Type | $T_{1/2}$ | σ ± Δσ |
|---|---|---|---|---|---|---|---|
| $^{190}$Ir | i(m1+g) | 11.78d | 0.334±0.069 | $^{82}$Br | i(m+g) | 35.30h | 1.94±0.19 |
| $^{189}$Ir | c | 13.2d | 49.8±5.4 | $^{75}$Se | c | 119.779d | 0.361±0.032 |
| $^{188}$Ir | i | 41.5h | 1.95±0.46 | $^{74}$As | i | 17.77d | 0.828±0.083 |
| $^{188}$Ir | c | 41.5h | 42.0±5.6 | $^{72}$Ga | c | 14.1h | 1.25±0.14 |
| $^{187}$Ir | c* | 10.5h | 46.1±6.7 | $^{72}$Ga | i | 14.1h | 0.92±0.13 |
| $^{186m}$Ir | i(m) | 1.90h | 23.1±4.3 | $^{72}$Zn | c | 46.5h | 0.331±0.040 |
| $^{186}$Ir | c | 16.64h | 16.2±1.4 | $^{59}$Fe | c | 44.472d | 0.377±0.033 |
| $^{185}$Ir | c | 14.4h | 23.9±3.2 | $^{46}$Sc | i(m+g) | 83.79d | 0.054±0.014 |
| $^{184}$Ir | c* | 3.09h | 25.3±2.2 | $^{7}$Be | i | 53.29d | 1.28±0.20 |

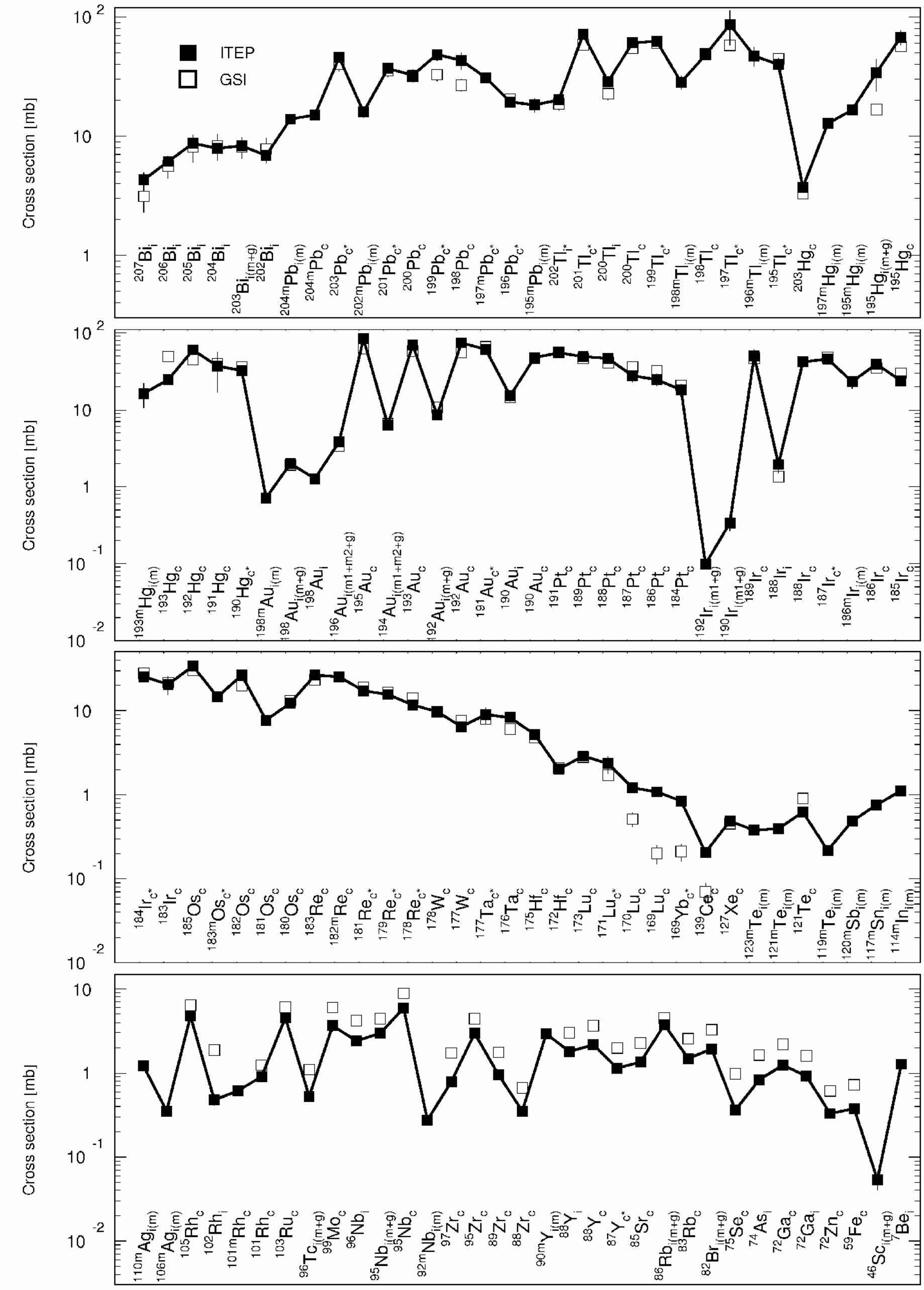

Fig. 18. Comparison of the nuclides production cross sections measured by the activation method at ITEP with estimated here cumulative yields from the independent cross sections by the inverse kinematics method measured at GSI for a $^{208}$Pb target irradiated with 500 MeV protons.